\documentclass[11pt, a4paper]{article}
\usepackage{jheppub}
\usepackage[T1]{fontenc}
\usepackage{mathtools}
\usepackage{cleveref}
\usepackage[nolist]{acronym}
\usepackage{subcaption}
\usepackage{tikz}
\usepackage{comment}
\usepackage{diagbox}

\crefname{figure}{figure}{figures} 

\graphicspath{{img/}}

\newcommand{\dd}{\text{d}}
\renewcommand{\i}{\ensuremath{\mathrm{i}}}
\newcommand{\eps}{\epsilon}
\newcommand{\re}{\mathrm{Re}}
\newcommand{\im}{\mathrm{Im}}
\newcommand{\avesum}[2]{\underset{#1}{\overset{#2}{\overline{\sum}}}}
\newcommand{\ds}{\mathrm{D}}

\newcommand{\neps}{{n_\eps}}
\newcommand{\nikv}{{n_v}}
\newcommand{\nfn}{{n_F}}

\newcommand{\lref}{\labelcref}
\newcommand{\incite}[1]{ref.~\cite{#1}}
\newcommand{\incites}[1]{refs.~\cite{#1}}

\definecolor{tab10blue}{HTML}{1f77b4}
\definecolor{tab10orange}{HTML}{ff7f0e}
\definecolor{tab10green}{HTML}{2ca02c}

\usepackage{tikz}
\usetikzlibrary{backgrounds,decorations.pathreplacing}

\tikzset{
    neuron/.style = {circle, minimum size=1.5em, line width=0},
    input neuron/.style = {neuron, fill=tab10green},
    output neuron/.style = {neuron, fill=tab10orange},
    hidden neuron/.style = {neuron, fill=tab10blue},
}

\newcommand{\layer}[4]{
    \foreach \name / \y in {1,...,#2}
    \node [#1] (#3-\name) at (#4,#2/2-\y) {};
}

\newcommand{\weights}[4]{
    \foreach \source in {1,...,#1}
    \foreach \dest in {1,...,#3}
    \draw [black!35,line width=0.8pt] (#2-\source) -- (#4-\dest);
}

\title{Learning Feynman integrals from differential equations with neural networks}

\author[a]{Francesco Calisto,}
\author[b]{Ryan Moodie,}
\author[c]{Simone Zoia}

\affiliation[a]{
    Ludwig-Maximilians-Universit\"at, Theresienstraße 39, 80333 M\"unchen, Germany
}
\affiliation[b]{
    Dipartimento di Fisica, Università di Torino, and INFN, Sezione di Torino, Via P.\ Giuria 1, I-10125 Torino, Italy
}
\affiliation[c]{
    CERN, Theoretical Physics Department, CH-1211 Geneva 23, Switzerland
}

\emailAdd{francesco.calisto@campus.lmu.de}
\emailAdd{ryaniain.moodie@unito.it}
\emailAdd{simone.zoia@cern.ch}

\abstract{
We perform an exploratory study of a new approach for evaluating Feynman integrals numerically.
We apply the recently-proposed framework of physics-informed deep learning to train neural networks to approximate the solution to the differential equations satisfied by the Feynman integrals.
This approach relies neither on a canonical form of the differential equations, which is often a bottleneck for the analytical techniques, nor on the availability of a large dataset, and after training yields essentially instantaneous evaluation times.
We provide a proof-of-concept implementation within the \texttt{PyTorch} framework, and apply it to a number of one- and two-loop examples, achieving a mean magnitude of relative difference of around 1\% at two loops in the physical phase space with network training times on the order of an hour on a laptop GPU.
}

\preprint{CERN-TH-2023-225}

\begin{document}

\maketitle
\flushbottom

\begin{acronym}
    \acro{DE}{differential equation}
    \acro{PDE}{partial differential equation}
    \acro{GELU}{gaussian error linear unit}
    \acro{MPL}{multiple polylogarithm}
    \acro{NN}{neural network}
    \acro{PIDL}{physics-informed deep learning}
    \acro{IBP}{integration-by-parts}
    \acro{MI}{master integral}
    \acro{ISP}{irreducible scalar product}
\end{acronym}

\section{Introduction}
\label{sec:Introduction}

The importance of Feynman integrals in theoretical physics can hardly be overstated.
They play a central role in quantum field theory, where they are necessary to obtain precise predictions for collider phenomenology and to unveil fundamental properties of the theory, but also find application in a growing range of other fields, such as gravitational waves, cosmology and statistical mechanics.
Their analytic expressions sport an intriguing `bestiary' of special functions, which has sparked the interest of mathematicians and has led to a fruitful cross-contamination between mathematics and physics.

The interest for Feynman integrals is therefore growing, and so is the effort in the search for techniques for evaluating them.
One of the most powerful approaches is the method of \acp{DE}~\cite{Barucchi:1973zm,Kotikov:1990kg,Kotikov:1991hm,Gehrmann:1999as,Bern:1993kr,Henn:2013pwa}, in which Feynman integrals are viewed as solutions to certain \acp{DE}.
Solving these \acp{DE} analytically in terms of well-understood classes of special functions guarantees efficient numerical evaluation and complete analytical control, but is not always feasible.
In such cases, one may resort to a numerical solution of the \acp{DE}.
In particular, the numerical solution through series expansions is gaining increasing interest after this technique was generalised from the univariate~\cite{Pozzorini:2005ff,Aglietti:2007as,Lee:2017qql,Lee:2018ojn,Bonciani:2018uvv,Fael:2021kyg,Fael:2022rgm} to the multivariate case~\cite{Liu:2017jxz,Moriello:2019yhu,Hidding:2020ytt,Liu:2021wks,Liu:2022chg,Armadillo:2022ugh}.
The most widely used fully numerical method is Monte Carlo integration with sector decomposition~\cite{Binoth:2000ps,Bogner:2007cr,Kaneko:2009qx,Borowka:2012yc,Borowka:2015mxa,Borowka:2017idc,Borowka:2018goh,Heinrich:2023til,Smirnov:2008py,Smirnov:2009pb,Smirnov:2013eza,Smirnov:2015mct,Smirnov:2021rhf}.
In addition to these, the long interest for Feynman integrals has led to a wide collection of other methods, such as Mellin-Barnes representation~\cite{Usyukina:1992jd,Usyukina:1993ch,Smirnov:1999gc,Tausk:1999vh,Czakon:2005rk,Smirnov:2009up,Gluza:2007rt,Belitsky:2022gba},
recurrence relations~\cite{Tarasov:1996br,Lee:2009dh,Lee:2012te},
symbolic integration~\cite{Brown:2008um,Panzer:2014caa},
loop-tree duality~\cite{Catani:2008xa,Runkel:2019yrs,Capatti:2019ypt},
Monte Carlo integration with tropical sampling~\cite{Borinsky:2020rqs,Borinsky:2023jdv},
and positivity constraints~\cite{Zeng:2023jek}.
We refer the interested readers to the comprehensive reviews~\cite{Smirnov:2012gma,Weinzierl:2022eaz}.
Recently, a few studies are beginning to explore the possibility of employing machine-learning methods in the evaluation of Feynman integrals.
In \incite{Winterhalder:2021ngy}, the authors use a \ac{NN} to optimise the choice of integration contour to optimise the Monte Carlo integration with sector decomposition.
Another approach consists in training a \ac{NN} to learn the primitive of an integral, then replacing the Monte Carlo integration with evaluations of the primitive~\cite{Maitre:2022xle,Cruz-Martinez:2023vgs}.
A compendium of machine-learning applications in particle physics is available in \incites{Feickert:2021ajf,Butter:2022rso,hepmllivingreview}.

In this paper we propose a new avenue: using machine learning to solve the \acp{DE} numerically.
We use deep \acp{NN} as universal function approximators~\cite{HORNIK1989359}, and train them to approximate the solution to the \acp{DE} by minimising a loss function that includes the \acp{DE} themselves along with a set of boundary values.
We build on the recently-proposed framework of \ac{PIDL}~\cite{raissi2017physicsI,raissi2017physicsII,raissi2019physics}.
In contrast to the traditional machine learning regime, where one trains on a large dataset, here we use a small dataset ---~the boundary values for the \acp{DE}~--- and strong physical constraints in the form of \acp{DE}.\footnote{\ac{PIDL} can be used also to solve integro-differential equations (see e.g.\ \incite{Dersy:2023job} and references therein).}
A preliminary application of this method to Feynman integrals was studied by one of the authors in his bachelor's thesis project~\cite{thesis-calisto}.
This approach is very flexible as it does not rely on a canonical form of the \acp{DE}~\cite{Henn:2013pwa}, unlike the analytical techniques, and after training yields essentially instantaneous evaluation times, unlike the numerical methods.
However, inherent precision limits mean this method cannot be competitive with analytical results, whenever they are available.~\footnote{For example, using 32-bit floating-point numbers implies a hard precision limit of about 8 digits.}

In addition to discussing our new approach, we provide a proof-of-concept implementation~\cite{repo,zenodo} within the \texttt{PyTorch} framework~\cite{paszke2019pytorch}, along with a number of one- and two-loop examples~\cite{models,zenodo}, some of them at the cutting edge of our current computational capabilities.
For these, we achieved a mean magnitude of relative difference of order 1\% at two loops in testing over the physical phase space, with \ac{NN} training times on the order of an hour on a laptop GPU.
These exploratory applications prove the feasibility of our approach, and allow us to identify the aspects which require further study in order to make this a fully developed alternative to current numerical methods.

This article is organised as follows.
In \cref{sec:DEs}, we review the method of \acp{DE} for calculating Feynman integrals.
In \cref{sec:ml}, we describe how we apply \ac{PIDL} to solve those \acp{DE}.
In \cref{sec:Examples}, we present example applications of our method to several integral families.
We conclude in \cref{sec:conc}.
\Cref{app:families} collects the definitions of the integral families considered in this work, 
while in \cref{app:plots} we gather figures detailing the training and testing statistics for all two-loop examples.

\section{Differential equations for Feynman integrals}
\label{sec:DEs}

In this section, we give a quick overview of the method of \acp{DE} for computing Feynman integrals~\cite{Barucchi:1973zm,Kotikov:1990kg,Kotikov:1991hm,Gehrmann:1999as,Bern:1993kr,Henn:2013pwa}.
In this approach, the problem of loop integration is traded for that of solving a system of first-order \acp{PDE} of the form
\begin{align} \label{eq:DEsIntro}
    \frac{\partial }{\partial v_i} \vec{F}(\vec{v}; \eps) = A_{v_i}(\vec{v} ; \eps) \cdot \vec{F}(\vec{v}; \eps) \,, \qquad \forall \, i = 1,\ldots, \nikv \,,
\end{align}
where $\vec{F} = (F_1,\ldots, F_\nfn)$ is an array of $\nfn$ Feynman integrals, $\vec{v} = (v_1,\ldots,v_{\nikv})$ are the $\nikv$ independent kinematic variables, and $\eps = (4-d)/2$ in $d$ spacetime dimensions.
The entries of the \emph{connection matrices} $A_{v_i}$ in \cref{eq:DEsIntro} are functions of $\vec{v}$ and $\eps$.
The \acp{DE} are complemented by (at least) one set of values $\vec{F}(\vec{v}_0; \eps)$ at an arbitrary phase-space point $\vec{v}_0$, which serves as the boundary condition.
We are interested in the Laurent expansion of the solution to the \acp{PDE} in \cref{eq:DEsIntro} around $\eps = 0$,
\begin{align} \label{eq:Fexp}
    \vec{F}(\vec{v};\eps) = \eps^{w^*} \sum_{w \ge 0} \eps^w F^{(w)}(\vec{v}) \,,
\end{align}
where $w^*$ is typically $- 2 \ell$ for an $\ell$-loop integral family.
We will truncate the expansion at some order $w_{\rm max}$ depending on the intended application.

In \cref{sec:DE-Box} we will show how the \acp{DE} can be derived for a given family of Feynman integrals using an explicit toy example.
The readers who are already familiar with this method may go directly to \cref{sec:DE-General}, which is devoted to a general characterisation of the \acp{DE} addressed in this work.
In \cref{sec:DE-Properties} we list a number of properties of the \acp{DE} and of their solutions which will be important for this work.

\subsection{Warm up: the massless box}
\label{sec:DE-Box}

In this section, we derive the \acp{DE} for the one-loop four-point ``massless box'' Feynman integrals.
A more pedagogical discussion of this example can be found e.g.\ in \incites{Henn:2014qga,Badger:2023eqz}.

An integral family is the set of all scalar integrals with a given propagator structure.
The integrals of the massless box family are defined as
\begin{align} \label{eq:box_def}
    \mathrm{I}_{\vec{a}}(s,t ; \eps) = \int \frac{\dd^d k}{\i \pi^{d/2}} \frac{ \mu^{4-d} }{D_1^{a_1} D_2^{a_2} D_3^{a_3} D_4^{a_4}} \,,
\end{align}
where $\mu$ is the dimensional regularisation scale, $\vec{a} = (a_1,a_2,a_3,a_4) \in \mathbb{Z}^4$, and $D_i$ are the inverse propagators associated with the graph in \cref{fig:box}:
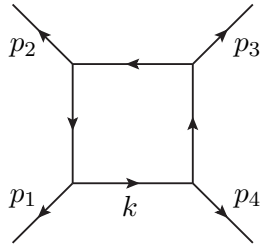
\begin{figure}
    \centering
    \begin{tikzpicture}
        \node at (0,0) {\includegraphics[scale=0.7]{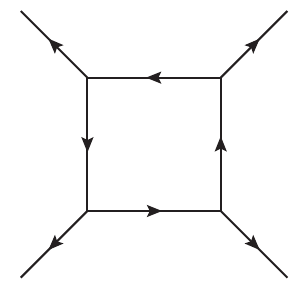}};
        \node at (-1.4,-1) {$p_1$};
        \node at (-1.4,1) {$p_2$};
        \node at (1.55,1) {$p_3$};
        \node at (1.55,-1) {$p_4$};
        \node at (0,-1.1) {$k$};
    \end{tikzpicture}
    \caption{
        Graph representing the propagator structure of the massless box integral family.
        The arrows denote the directions of the momenta.
    }
    \label{fig:box}
\end{figure}
\begin{align}
    \begin{alignedat}{2}
        D_1 & = - k^2 - \i 0^+ \,, \qquad \qquad && D_3 = -(k+p_1+p_2)^2 - \i 0^+ \,, \\
        D_2  & = - (k+p_1)^2 - \i 0^+ \,, \qquad \qquad && D_4 = - (k-p_4)^2 - \i 0^+ \,,
    \end{alignedat}
\end{align}
with $0^+$ a positive infinitesimal.
We set $\mu=1$; the dependence on it can be recovered from dimensional analysis.
The external momenta $p_i$ are taken to be outgoing, and satisfy momentum conservation and on-shell conditions:
\begin{align}
    \sum_{i=1}^4 p_i = 0 \,, \qquad \qquad p_i^2 = 0 \quad \forall \, i = 1,\ldots, 4 \,.
\end{align}
We choose the two independent Lorentz invariants as $s = (p_1+p_2)^2$ and $t = (p_2+p_3)^2$.

Within each family only finitely many integrals are linearly independent.
They are called \acp{MI}, and constitute a basis of the family.
Any integral of the family can thus be ``reduced'' to a linear combination of \acp{MI}.
The coefficients are rational functions of the kinematic invariants and $\epsilon$.\footnote{One may introduce more complicated factors, such as square roots, in the definition of the \acp{MI}.}
The standard way to determine the \acp{MI} and reduce to \acp{MI} is to solve systems of \ac{IBP} relations~\cite{Tkachov:1981wb,Chetyrkin:1981qh} ---~linear relations among the integrals of a family~--- using the Laporta algorithm~\cite{Laporta:2000dsw}.
We generate the required \ac{IBP} relations using \texttt{LiteRed}~\cite{Lee:2012cn}, and solve them using \texttt{FiniteFlow}~\cite{Peraro:2019svx}.
The box family has three \acp{MI}.
Following the Laporta algorithm, they are chosen as
\begin{align} \label{eq:box-noncanMI}
    \vec{F}(s,t ; \epsilon) = \begin{pmatrix}
        \mathrm{I}_{0,1,0,1}(s,t ; \eps) \,, &
        \mathrm{I}_{1,0,1,0}(s,t ; \eps) \,, &
        \mathrm{I}_{1,1,1,1}(s,t ; \eps)
    \end{pmatrix}^{\top} \,.
\end{align}
We stress that the choice of \acp{MI} is arbitrary, provided they are linearly independent.

The derivatives of the \acp{MI} with respect to the kinematic invariants can be expressed as linear combinations of scalar integrals in the same family, and can thus be \ac{IBP}-reduced to \acp{MI} themselves.
As a result, we obtain a system of first-order \acp{PDE}:
\begin{align} \label{eq:box-noncanDE}
    \frac{\partial}{\partial s} \vec{F}(s,t ; \epsilon)  =
    \begin{pmatrix}
        0 & 0 & 0 \\
        0 & -\frac{\eps}{s} & 0 \\
        \frac{2 (2 \eps-1)}{s t (s + t)} &  \frac{2 (1 - 2 \eps)}{s^2 (s + t)} & -\frac{s + t + \eps \, t}{s (s + t)} \\
    \end{pmatrix} \cdot \vec{F}(s,t ; \epsilon) \,,
\end{align}
and similarly for $t$.

We are interested in the Laurent expansion of the solution around $\eps = 0$,
\begin{align} \label{eq:Fexp-box}
    \vec{F}(s,t; \eps) = \frac{1}{\eps^2} \sum_{k \ge 0} \eps^k \, \vec{F}^{(k)}(s,t) \,,
\end{align}
truncated at a certain order in $\eps$.
To this end, the standard approach consists in looking for a \ac{MI} basis such that the \acp{DE} simplify.
For example, choosing
\begin{align} \label{eq:box-canMIs}
    \vec{F}(s,t ; \epsilon) = \begin{pmatrix}
        s \, t \, \mathrm{I}_{1,1,1,1}(s,t ; \eps) \,, &
        s \, \mathrm{I}_{1,1,1,0}(s,t ; \eps) \,, &
        t \, \mathrm{I}_{1,1,0,1}(s,t ; \eps)
    \end{pmatrix}^{\top}
\end{align}
as \acp{MI} for the massless box family leads to \acp{DE} of the form
\begin{align} \label{eq:canDE-box}
    \frac{\partial}{\partial s} \vec{F}(s,t ; \epsilon)  =
    \eps \, A_s(s,t)  \cdot \vec{F}(s,t ; \epsilon) \,,
\end{align}
with
\begin{align}
    A_s(s,t) = \begin{pmatrix}
        \frac{1}{s + t}-\frac{1}{s}  & \frac{2}{s + t}-\frac{2}{s}  & \frac{2}{s + t} \\
        0 &  -\frac{1}{s} & 0 \\
        0 & 0 & 0 \\
    \end{pmatrix} \,,
\end{align}
and similarly for $t$.
The \acp{DE} in \cref{eq:canDE-box} are in the so-called \emph{canonical form}~\cite{Henn:2013pwa}: the connection matrices depend on $\eps$ only through an overall factor and have at most simple poles at all singular loci.
The factorisation of $\eps$ allows for an iterative solution of the \acp{DE} order by order in $\eps$.
Plugging the Laurent expansion of \cref{eq:Fexp-box} into \cref{eq:canDE-box} in fact gives the recursion
\begin{align}
    \frac{\partial}{\partial s} \vec{F}^{(k)}(s,t)  =
    A_s(s,t)  \cdot \vec{F}^{(k-1)}(s,t)  \,,
\end{align}
for $k\ge 1$, starting from $\vec{F}^{(0)}(s,t)$ which is constant.
If the connection matrices are rational functions, the solution can be written algorithmically in terms of \acp{MPL} up to any order in $\eps$.\footnote{In the presence of square roots which cannot be rationalised by a change of variables it may still be possible to express the solution in terms of \acp{MPL} (see e.g.\ \cite{Heller:2019gkq,Canko:2020ylt,Kardos:2022tpo}), but this is in general not possible beyond one loop~\cite{Duhr:2020gdd}.}
The boundary values can be determined by evaluating the \acp{MI} numerically, or by imposing physical consistency conditions (see \incites{Henn:2014qga,Badger:2023eqz}).
For example, for the first \ac{MI} in \cref{eq:box-canMIs} we obtain
\begin{align} \label{eq:box-sol}
    \begin{aligned}
        F_1(s,t;\eps) = \, & \frac{\mathrm{e}^{-\eps \gamma_{\rm E}} (-s)^{-\eps}}{\eps^2} \biggl\{ 4 - 2 \eps \log\left(\frac{t}{s}\right) - \frac{4}{3} \pi^2 \eps^2 + \eps^3 \biggl[ 2 \, \mathrm{Li}_3\left(-\frac{t}{s}\right) \\
        & - 2 \log\left(\frac{t}{s}\right) \mathrm{Li}_2\left(-\frac{t}{s}\right)
        - \log^2\left(\frac{t}{s}\right) \log\left(\frac{s+t}{t}\right)  + \frac{1}{3} \log^3\left(\frac{t}{s}\right) \\
        & + \frac{7}{6} \pi^2 \log\left(\frac{t}{s}\right) - \pi^2  \log\left(\frac{s+t}{s}\right) - \frac{34}{3} \zeta_3 \biggr] + \mathcal{O}\left(\eps^4 \right) \biggr\}\,,
    \end{aligned}
\end{align}
where ${\rm Li}_n$ is the weight-$n$ classical polylogarithm, $\zeta_n$ is the Riemann zeta constant, and $\gamma_{\rm E}$ is the Euler-Mascheroni constant.\footnote{The expression in \cref{eq:box-sol} is well defined in the Euclidean region ($s<0 \land t<0$).
The analytic continuation to the physical region ($s>0 \land t<0$) is obtained by adding a small positive imaginary part to both $s$ and $t$. 
For example, this amounts to replacing $\log(t/s)$ with $\log(-t/s)+\i \pi$.}

\subsection{General form of the differential equations}
\label{sec:DE-General}

In the previous section, we have seen that finding a \ac{MI} basis that satisfies canonical \acp{DE} simplifies the solution substantially.
This is however a difficult endeavour, for which no general algorithm exists.
What is more, special functions that are more complicated than \acp{MPL} may be required to express the solution analytically (see e.g.\ \incite{Bourjaily:2022bwx} for a recent review).
The notion of ``canonical'' \acp{DE} in such cases is still object of ongoing research, and the mathematical technology to handle the relevant special functions is far from being ripe.

Our approach does not rely on the canonical form of the \acp{DE}, and is insensitive to the class of special functions involved in the solution.
We will make only the following mild assumptions about the connection matrices $A_{v_i}(\vec{v};\eps)$ in \cref{eq:DEsIntro}.
We will assume that:
\begin{enumerate}
    \item they are \emph{rational} functions of the kinematic variables $\vec{v}$ and $\eps$;
    \item they are \emph{finite} at $\eps=0$, and can thus be expanded as
        \begin{align} \label{eq:DEs-exp}
            A_{v_i}(\vec{v};\eps) = \sum_{k=0}^{k_{\rm max}} \eps^k A^{(k)}_{v_i}(\vec{v})  \,,
        \end{align}
        where $k_{\rm max}$ is either a positive integer or infinity.
\end{enumerate}
We stress that both these assumptions are in principle not necessary for our method, but are useful simplifications.

The first condition is easy to achieve.
Square roots or more complicated functions in fact appear in the connection matrices only if they are present in the definition of the \acp{MI}.
If the latter contain only rational functions, as in \cref{eq:box-canMIs} for the massless box family, the connection matrices are guaranteed to be rational.
This condition implies that the connection matrices are real-valued for real kinematic variables $\vec{v}$, which allows us to decouple the real and imaginary parts of the \acp{MI}.
The latter satisfy the same \acp{DE},
\begin{align}
    & \frac{\partial}{\partial v_i} \re\left[ \vec{F}(\vec{v};\eps) \right] = A_{v_i}(\vec{v}; \eps) \cdot \re\left[ \vec{F}(\vec{v};\eps) \right] \,, \\
    & \frac{\partial}{\partial v_i} \im\left[ \vec{F}(\vec{v};\eps) \right] = A_{v_i}(\vec{v}; \eps) \cdot \im\left[ \vec{F}(\vec{v};\eps) \right] \,,
\end{align}
though with different boundary values.
Furthermore, this assumption allows us to avoid any operations with complex numbers, making the implementation simpler and more efficient.

The construction of a \ac{MI} basis which satisfies the second condition is less trivial (see e.g.\ \incites{Lee:2019wwn,Dubovyk:2022frj}\footnote{We are grateful to Christoph Dlapa and Johann Usovitsch for pointing out these references.}), but is substantially simpler than obtaining a canonical form.
For the most complicated family computed in this paper, for instance, it simply amounted to choosing the \acp{MI} according to the algorithm of \incite{Smirnov:2020quc,Usovitsch:2020jrk}, and normalising some of them by suitable factors of $\eps$.
We will discuss how we achieved this for each of the examples in \cref{sec:Examples}.
The advantage of this form of the connection matrices is that it leads to a recursive structure of the solution where the coefficients of the $\eps$-expansion of the \acp{MI} in \cref{eq:Fexp} are coupled by the \acp{DE} only to themselves and to lower-order coefficients:
\begin{align}
    \frac{\partial}{\partial v_i} \vec{F}^{(w)}(\vec{v}) = \sum_{k=0}^{\mathrm{min}(w,k_{\rm max})} A^{(k)}_{v_i}(\vec{v}) \cdot \vec{F}^{(w-k)}(\vec{v}) \,.
\end{align}
While our method works in principle for any $k_{\rm max}$, low values are preferable.
It is therefore worth putting some effort into reducing $k_{\rm max}$.
Moreover, since we are interested in the solution only up to a certain order, say $w_{\rm max}$, the terms of the connection matrices of order greater than $w_{\rm max}$ do not contribute.

\smallskip

In conclusion, we will solve systems of \acp{PDE} of the form
\begin{align} \label{eq:DEs-finalform}
    \frac{\partial}{\partial v_i} \vec{G}^{(w)}(\vec{v}) = \sum_{k=0}^{\mathrm{min}(w,k_{\rm max})} A^{(k)}_{v_i}(\vec{v}) \cdot \vec{G}^{(w-k)}(\vec{v}) \,, \qquad \forall \ i=1,\ldots,\nikv\,,
\end{align}
from $w=0$ up to a given order, complemented by (at least) one set of boundary values $\vec{G}^{(w)}(\vec{v}_0)$ obtained by evaluating numerically the \acp{MI}.
Here, $\vec{G}^{(w)}(\vec{v})$ is either the real or the imaginary part of the \ac{MI} coefficients $\vec{F}^{(w)}(\vec{v})$, and the entries of the connection matrices $A^{(k)}_{v_i}(\vec{v})$ are rational functions of the kinematic variables $\vec{v}$.


\subsection{Properties of the differential equations}
\label{sec:DE-Properties}

In this section, we collect a number of properties of the \acp{MI} and of the corresponding \acp{DE}, which will play a role in our machine learning approach to the solution.

\paragraph{Homogeneity}
We choose the kinematic variables $\vec{v}$ as either scalar products $(p_i+p_j)^2$ or squared masses $m_i^2$.
The \acp{MI} are then homogeneous functions of $\vec{v}$, i.e.\ they satisfy the scaling condition
\begin{align} \label{eq:scaling}
    {F}_i(\alpha \, \vec{v}; \eps ) = \alpha^{\lambda_i} {F}_i(\vec{v} ; \eps)  \,,
\end{align}
for any positive factor $\alpha$, where the scaling dimensions $\lambda_i$ can be determined by dimensional analysis.
The scaling in \cref{eq:scaling} can be used to set one variable to a constant.
We can for instance ``freeze'' the last variable $v_{\nikv}$ to a constant $c_{\nikv}$ by choosing $\alpha = c_{\nikv}/v_{\nikv}$ in \cref{eq:scaling}, obtaining
\begin{align} \label{eq:F-frozen}
    {F}_i(\vec{v} ; \eps ) = \left(\frac{v_{\nikv}}{c_{\nikv}}\right)^{\lambda_i}  \, {F}_i\left(c_{\nikv} \frac{v_1}{v_{\nikv}}, \ldots, c_{\nikv} \frac{v_{\nikv-1}}{v_{\nikv}} , c_{\nikv} ; \eps \right) \,.
\end{align}
This allows us to treat the integrals as functions of $\nikv-1$ ratios.
The choice of which variable to freeze and of which value to fix it to both have an impact on the \ac{NN} fit.
Similarly, it is sometimes convenient to choose the \acp{MI} so that their scaling dimensions $\lambda_i$ in \cref{eq:scaling} are all zero.
These choices in fact affect the range of scales in the inputs and the target functions within the chosen kinematic region.
Stochastic gradient descent algorithms are in fact highly sensitive to the range of scales involved in the problem, and train optimally when all scales are of the same order.
These choices should therefore be done so as to minimise ---~as much as possible~--- the range of scales in the problem.
This is part of the hyper-parameter tuning, which in this study we performed by hand.
We will discuss these aspects in \cref{sec:ml}.

\paragraph{Singularities}
The denominators of the connection matrices encode the singularity structure of the solution to the \acp{DE}.
In other words, the solution to the \acp{DE} may be singular where any of the denominator factors of the connection matrices vanishes.
We distinguish between spurious and physical singularities.
On a \emph{spurious singularity}, the connection matrices are singular while the solution stays finite.
On a \emph{physical singularity}, instead, also the solution is singular.
For the massless box integrals discussed in \cref{sec:DE-Box}, for example, the denominator factors of the \acp{DE} are $\{s,t,s+t\}$.
Of these, $s=0$ and $t=0$ are physical singularities, while $s+t=0$ is spurious.

\paragraph{Multivaluedness}
Feynman integrals, and hence the special functions appearing in them, are multivalued functions.
Care must be taken that they are evaluated on the correct branch.
The expression for the box integral in \cref{eq:box-sol} is for instance well-defined in the Euclidean region ($s<0 \land t<0$).
Some work is required to analytically continue it to the physical region of interest for phenomenology ($s>0 \land t<0$).
We sidestep this problem by working always within a given kinematic region, never crossing its borders into another one.
In practice, this means that we choose boundary points in the chosen region, and fit the solution to the DEs in that region only.
This way the boundary values instruct the \ac{NN} which branch of the solution to fit, and no explicit analytic continuation is required.

\section{Deep learning Feynman integrals}
\label{sec:ml}

In this section, we detail our application of machine learning techniques to solving Feynman integrals from their \acp{DE}.
Note that all hyperparameters are selected manually without extensive optimisation, this being an exploratory study.

In brief, \acp{NN} are trained by minimising a heuristic that we call the \emph{loss function}, typically using a variant of stochastic gradient descent~\cite{Goodfellow-et-al-2016}.
By evaluating the loss function and then considering its derivatives with respect to the parameters of the \ac{NN}, we can adjust those parameters to reduce the value of the loss function.
Repeating this iteratively, we converge on a minimum of the loss function.
The derivatives of the loss function are generally computed using an algorithmic method called automatic differentiation~\cite{doi:10.1137/1.9780898717761}.

This machinery can be used for regression analysis, where the \ac{NN} becomes a function approximator.
Say we wish to emulate a target function $g$ with a \ac{NN} surrogate $h$.
If we have a large dataset of $N$ values of the target at points $x^{(i)}$,
\begin{align}
    \ds \coloneqq \left\{\left(x^{(i)}, g\bigl(x^{(i)}\bigr)\right) \,\Bigl|\, i=1,\ldots,N\right\} \,,
\end{align}
we can use a loss function $L(\ds, \theta)$ that compares this target distribution to the output of our \ac{NN}, where $\theta$ are the parameters (weights and biases) of the \ac{NN}.
What constitutes a large $N$ depends on the number of input dimensions and the complexity of the target distribution.
A common choice is the mean squared error loss function,
\begin{align}
    L(\ds, \theta) &= \frac{1}{N} \sum_{i=1}^N \Bigl[ h\bigl(x^{(i)}; \theta\bigr) - g\bigl(x^{(i)}\bigr)\Bigr]^2 \,.
\end{align}

For integral families of interest, we do not have access to a large dataset because generating sample points with existing evaluation methods is prohibitively slow, so this approach is unviable.
However, we do know the \acp{DE} \lref{eq:DEs-finalform} and can cheaply numerically sample their connection matrices.
Therefore, we can instead train a \ac{NN} to satisfy the \acp{DE} by including in our loss function a
term which minimises the difference between the two sides of the \acp{DE} as estimated by the \ac{NN}.
To fully determine the solution, we also specify at least one set of boundary values as another constraint term in our loss function.
This approach comes under the paradigm of \acf{PIDL}~\cite{raissi2017physicsI,raissi2017physicsII,raissi2019physics}; we refer the interested reader to the tutorials~\cite{navarro2023solving,baty2023solving} and recent reviews of related literature~\cite{Karniadakis2021,cuomo2022scientific,hao2023physicsinformed,faroughi2023physicsguided}. 

Our application of \ac{PIDL} is however somewhat peculiar.
Usually, in the machine learning community, \ac{PIDL} is used in architectures which blend data-driven and model-driven (i.e., physics-informed) components. 
The role of the latter is to constrain the network to physically sensible results (for instance, by imposing energy conservation), while the remaining part of the learning is still performed using a data-driven loss function. 
In that setting, \ac{PIDL} contributes to the ``physics for machine learning'' cause. 
On the contrary, we limit the data-driven part to a few boundary conditions and leave the bulk of learning to be model-driven, as the \acp{DE} themselves represent, perturbatively, an exact solution to the problem. 
Ours is therefore an instance of the ``machine learning for physics''~cause.

This section is organised as follows.
In \cref{sec:arch}, we discuss the architecture of our model.
In \cref{sec:data}, we detail the datasets and the subtleties of their creation.
In \cref{sec:loss}, we define our loss function.
In \cref{sec:train}, we describe the training procedure.
Finally, in \cref{sec:error}, we cover the uncertainty and error analysis.

\subsection{Model architecture}
\label{sec:arch}

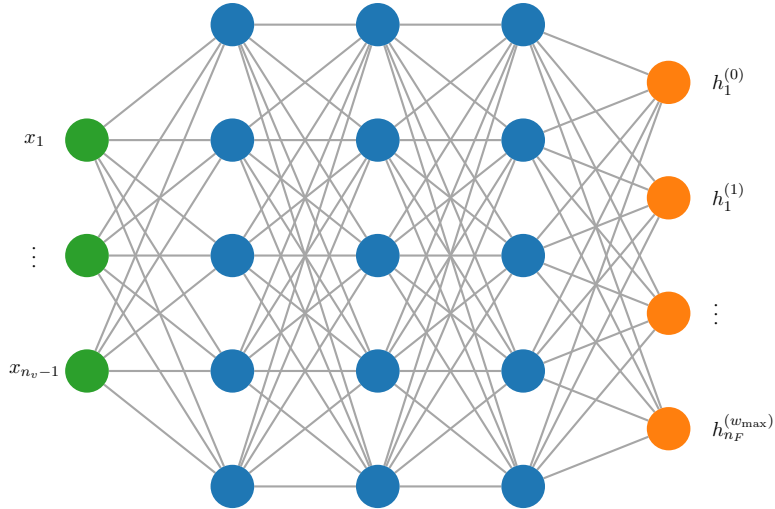
\begin{figure}
    \begin{center}
        \begin{tikzpicture}[x=5em, y=4em]
            \newcommand{\nin}{3};
            \newcommand{\nha}{5};
            \newcommand{\nhb}{5};
            \newcommand{\nhc}{5};
            \newcommand{\no}{4};

            \newcommand{\rvdots}{\rotatebox{90}{$\cdots$}};

            \layer{input neuron}{\nin}{I}{0}
            \layer{hidden neuron}{\nha}{HA}{1}
            \layer{hidden neuron}{\nhb}{HB}{2}
            \layer{hidden neuron}{\nhc}{HC}{3}
            \layer{output neuron}{\no}{O}{4}

            \begin{scope}[on background layer]
                \weights{\nin}{I}{\nha}{HA}
                \weights{\nha}{HA}{\nhb}{HB}
                \weights{\nhb}{HB}{\nhc}{HC}
                \weights{\nhc}{HC}{\no}{O}
            \end{scope}

            \node [scale=0.7] at (-0.36,0.5) {$x_1$};
            \node [scale=0.7] at (-0.36,-0.5) {\rvdots};
            \node [scale=0.7] at (-0.36,-1.5) {$x_{\nikv-1}$};
            \node [scale=0.7, anchor=west] at (4.25,1.01) {$h_1^{(0)} $};
            \node [scale=0.7, anchor=west] at (4.25,0.01) {$h_1^{(1)} $};
            \node [scale=0.7, anchor=west] at (4.25,-0.99) {\rvdots};
            \node [scale=0.7, anchor=west] at (4.25,-1.99) {$h_\nfn^{(w_{\rm max})}$};

        \end{tikzpicture}
    \end{center}
    \caption{
        Architecture of the \ac{NN} with outputs $h_i^{(w)}(\vec{x})$ to fit the \ac{MI} coefficients $g_i^{(w)}(\vec{x})$ as described in \cref{sec:arch}.
        The input nodes are shown in green, the hidden layer nodes in blue, and the output nodes in orange.
        In our applications, the number of output nodes is much larger than that of the input nodes.
    }
    \label{fig:NN}
\end{figure}

We approximate the \ac{MI} coefficients $\vec{F}^{(w)}(\vec v)$ for $w=0,\ldots, w_{\rm max}$ with a model comprising a pair of fully-connected feedforward \acp{NN}.
Each \ac{NN} is a real-valued function, with one for the real and one for the imaginary parts of the \ac{MI} coefficients, which we collectively refer to as $\vec{G}^{(w)}(\vec v)$ as in \cref{eq:DEs-finalform}.

The \acp{NN} have $\nikv-1$ inputs $\vec x \coloneqq (x_1, \ldots, x_{\nikv-1})$, which are related to a subset of the $\nikv$ kinematic variables $\vec v$.
We make use of the homogeneity of the solution (see \cref{sec:DE-Properties}) to drop the dependence on one of the variables, as decreasing the input dimensionality reduces the complexity of the problem.
The choice of which variable to freeze is made considering the kinematics of the particular family;
for instance, natural choices include $s_{12} = (p_1+p_2)^2$ for a massless system with incoming momenta $p_1$ and $p_2$, or the (squared) mass in a system with an internal mass.
If we fix the last variable $v_{\nikv}$ to a constant $c_{\nikv}$ and define the variables
\begin{align}
    \label{eq:kinematic-scaling}
    x_i \coloneqq c_{\nikv} \frac{v_i}{v_{\nikv}} \qquad \forall \, i = 1,\ldots,\nikv-1 \,,
\end{align}
from \cref{eq:F-frozen} ---~which holds separately at each order in $\eps$, and for both real and imaginary parts~--- it follows that
\begin{align}
    G_i^{(w)}(\vec{v}) =  \left(\frac{v_{\nikv}}{c_{\nikv}}\right)^{\lambda_i} g_i^{(w)}\left(x_1,\ldots,x_{\nikv-1}\right) \,,
\end{align}
where $g_i^{(w)}$ is a function of $\nikv-1$ variables $\vec{x}$,
\begin{align}
    g_i^{(w)}(\vec{x}) \coloneqq G_i^{(w)}(x_1, \ldots, x_{\nikv-1}, c_{\nikv} ) \,.
\end{align}
The sets of functions $\vec{g}^{(w)}(\vec{x}) \coloneqq (g_1^{(w)}, \ldots, g_\nfn^{(w)})$ for $w=0,\ldots, w_{\rm max}$ then satisfy \acp{DE} of the same form as \cref{eq:DEs-finalform} for the \ac{MI} coefficients,
\begin{align} \label{eq:DEs-finalform-f}
    \frac{\partial}{\partial x_i} \vec{g}^{(w)}(\vec{x}) =
    \sum_{k=0}^{\mathrm{min}(w,k_{\rm max})} A^{(k)}_{x_i}(\vec{x}) \cdot \vec{g}^{(w-k)}(\vec{x}) \,,
\end{align}
for $i=1,\ldots,\nikv-1$ and $w=0,\ldots,w_{\rm max}$,
with connection matrices $A^{(k)}_{x_i}(\vec{x})$ which can be derived from those in \cref{eq:DEs-finalform} through the chain rule.

Our \acp{NN} are trained to fit the sets of functions $\vec{g}^{(w)}$ for $w$ from $0$ to $w_{\rm max}$, that is
\begin{align}
    \neps \coloneqq w_{\rm max}+1
\end{align}
orders in $\eps$.
The outputs of a \ac{NN} correspond to the functions $g_i^{(w)}(\vec x)$.
Each \ac{NN} thus has $\nfn \, \neps$ outputs,
which are typically many more than the inputs $\vec{x}$. 
We recall that $n_F$ is the number of \acp{MI}.

For the integral families we considered, we find that three or four hidden layers are sufficient.
Validation performance saturates with square (constant width) configurations with a width of around double the number of outputs for three layers or similar to the number of outputs for four layers.
The architecture of the \ac{NN} is sketched in \cref{fig:NN}.
The choice of activation functions is critical in \ac{PIDL}~\cite{wang2023learning};
we use \ac{GELU} activation functions\footnote{In \ac{PIDL}, the activation functions must have nonzero and continuous second-order derivatives in order for stochastic gradient descent to work (see~\cref{sec:loss}).
We find \ac{GELU} performs similarly or favourably to the canonical choice of the hyperbolic tangent function in our examples.}~\cite{hendrycks2023gaussian} on the hidden layers, with a linear output layer.
We use 32 bit floating point numbers, finding this numerical precision to not be a limiting factor of the training performance.

\subsection{Datasets}
\label{sec:data}

The training dataset $\ds$ has two subsets: a \ac{DE} dataset $\ds_{\rm DE}$ and a boundary dataset $\ds_{\rm b}$.
The \ac{DE} dataset comprises a set of input points and the corresponding values of the connection matrices:
\begin{align}
    \label{eq:de-dataset}
    \ds_{\rm DE}\coloneqq \left\{\left(\vec{x}^{(i)}, A_{x_j}^{(k)}\bigl(\vec{x}^{(i)}\bigr) \right) \, \Bigl| \, i = 1,\ldots \right\} \,.
\end{align}
Here, by $A_{x_j}^{(k)}\bigl(\vec{x}^{(i)}\bigr)$ we mean the set of connection matrices for all variables ($j = 1,\ldots,\nikv-1$) and at all the relevant orders in $\eps$ ($k=0,\ldots,k_{\rm max}$), evaluated at $\vec{x}^{(i)}$.
The \ac{DE} dataset is generated dynamically by random sampling at each iteration of the training algorithm (see \cref{sec:train}), while all other datasets are static and pre-computed.
The boundary dataset contains input points with values of the solution,
\begin{align}
    \label{eq:bc-dataset}
    \ds_{\rm b}\coloneqq \left\{\left(\vec{x}^{(i)}, \vec{g}^{(w)}\bigl(\vec{x}^{(i)}\bigr) \right) \, \Bigl| \, i = 1,\ldots \right\} \,,
\end{align}
where $\vec{g}^{(w)}\bigl(\vec{x}^{(i)}\bigr)$ includes all required orders in $\eps$ ($w=0,\ldots,w_{\rm max}$).
There is also another dataset of this form for testing.
64 bit floating point numbers are used to generate and store all datasets, with the exception of solution values computed with arbitrary-precision arithmetic by \texttt{AMFlow}~\cite{Liu:2017jxz,Liu:2021wks,Liu:2022chg} as noted in \cref{sec:Examples}.

\smallskip

As we anticipated in \cref{sec:DE-Properties}, we target a single phase-space region, where the solution is single-valued, so that no analytic continuation needs to be performed.
The phase-space points of all datasets are thus chosen within said region, and the fit is valid only there.
The boundaries of the kinematic regions are physical singularities (both connection matrices and solution are singular), while the bulk of the phase space can be crossed by both physical and spurious singularities (the connection matrices are singular but the solution is finite).
When sampling points for all datasets, we must thus place \emph{cuts} around the physical singularities.
Furthermore, for the \ac{DE} dataset only, we must also cut the spurious poles to avoid divergences in the numerical evaluation of the connection matrices.
We stress however that the \acp{NN} can be evaluated also in the regions surrounding the spurious singularities excluded from the \ac{DE} dataset, and actually exhibit good accuracy there as well.
We choose the cuts to be as small as possible while regulating the singular behaviour to minimise the absence of model exposure to these regions during training.

More concretely, we parametrise the phase space of the physical scattering region of interest in terms of energies and angles, then generate the points of the input space $\vec{x}$ for \ac{DE} and testing datasets by random uniform sampling of the parametrisation.
In addition, we veto points that do not pass the appropriate singularity cuts.
This provides a simple sampling method for this exploratory study.
While we did not experiment with different phase-space distributions, we expect this to have an impact on the model performance.
A given phase-space generator may in fact sample more densely where the target functions vary the most, and this should be reflected in the training.
For this reason, it is important to use the same phase-space generator in the training as in the intended application, e.g.\ the Monte Carlo integration to compute a cross section.
Firstly, this will provide the optimal training input distribution.
Secondly, the uncertainty estimates discussed below are tied to the input distribution used during the training, and therefore only make sense in an application if it employs the same distribution.

\smallskip

Stochastic gradient descent algorithms are highly sensitive to the range of magnitudes involved in the problem, training optimally when all scales are of the same order.
In particular, \emph{large scale variations} in the distributions of the inputs and outputs can be problematic.
We select from two methods to ameliorate the output distribution, finding that the better option depends on the integral family:
\begin{itemize}
    \item constructing integrals with null scaling dimensions (see \cref{eq:scaling}), including tuning the choice of prefactor used to normalise the integrals;
    \item using a judicious choice of the kinematic scaling $c_{\nikv}$ in \cref{eq:kinematic-scaling} for integrals with non-zero scaling dimensions.
\end{itemize}
The choice of $c_{\nikv}$ can also be exploited to adjust the input distribution and shift the spurious surfaces such that the spurious cuts exclude less of the physical region.
Cumulatively, these choices can have a significant impact on training performance and must be fine tuned.

\smallskip

While it is sufficient to specify a single boundary point, we find that training performance can be improved by including additional points in the boundary dataset.
The optimal configuration depends on the specific system, but we note some trends.
For small boundary datasets, we find that points on the phase-space boundaries (including physical and spurious cuts) are more important than those in the interior, while having a mix of both becomes beneficial as the dataset size increases.
If using random sampling of the interior, the significance of points on the boundary decreases with dataset size as more interior points randomly land near the boundaries.
Performance generally increases with the total number of points, although the improvement diminishes as the size becomes very large.
It is vital to choose points that well represent the scattering region, including coverage of the boundaries; we find that large randomly generated sets with many overlapping points can even perform worse than smaller ones with a more even distribution, for instance.

\subsection{Loss function}
\label{sec:loss}

We define the loss function $L(\ds,\theta)$ for a \ac{NN} with outputs $\vec{h}^{(w)}(\vec{x}; \theta)$ fitting a target solution $\vec{g}^{(w)}(\vec{x})$ for inputs $\vec{x}$.
We write it as a sum of terms for the \acp{DE} ($L_{\rm DE}$) and for the boundary values ($L_{\rm b}$), using mean squared error for each term.
Using the short-hand
\begin{align}
    \avesum{i \in A}{} \coloneqq \frac{1}{|A|} \sum_{i \in A} \ ,
\end{align}
with $|A|$ the cardinality of the set $A$,
the \ac{DE} terms of the loss function are
\begin{align} \label{eq:L-DE}
    \begin{aligned}
        & L_{\rm DE}(\ds_{\rm DE},\theta)=  \\
        & \qquad \avesum{\vec{x}^{(i)} \in \ds_{\rm DE}}{} \, \avesum{j=1}{\nfn} \avesum{l=1}{\nikv-1} \avesum{w=0}{w_{\rm max}} \Biggl[
            \partial_{x_l} h_j^{(w)}\bigl(\vec{x}^{(i)}; \theta\bigr) -
            \sum_{k=0}^{\min(w,k_{\rm max})} \sum_{r=1}^{\nfn} A_{x_l,jr}^{(k)}\bigl(\vec{x}^{(i)}\bigr) \, h_r^{(w-k)}\bigl(\vec{x}^{(i)}; \theta\bigr)
            \Biggr]^2 \,,
    \end{aligned}
\end{align}
while the boundary terms are given by
\begin{align} \label{eq:L-b}
    L_{\rm b}(\ds_{\rm b},\theta) &=
    \avesum{\vec{x}^{(i)}\in \ds_{\rm b}}{} \, \avesum{j=1}{\nfn} \avesum{w=0}{w_{\rm max}}  \left[
        h_j^{(w)}\bigl(\vec{x}^{(i)}; \theta\bigr) -
        g_j^{(w)}\bigl(\vec{x}^{(i)}\bigr)
        \right]^2 \,.
\end{align}
We sum them to obtain the full loss function,\footnote{
    The relative weight of the \ac{DE} and boundary terms can be tuned by multiplying either of them by a parameter.
    This may be useful in case the solutions and their derivatives have very different scales.
    We did not find this to be necessary in the examples considered here.
}
\begin{align} \label{eq:L-full}
    L(\ds,\theta) = L_{\rm DE}(\ds_{\rm DE},\theta) + L_{\rm b}(\ds_{\rm b},\theta) \,.
\end{align}
We recall that $\ds \coloneqq \ds_{\rm DE} \cup \ds_{\rm b}$ is the training dataset defined in \cref{eq:de-dataset,eq:bc-dataset}, $\theta$ are the parameters of the \ac{NN}, and $A^{(k)}_{x_l}$ are the connection matrices in \cref{eq:DEs-finalform-f}.

To obtain the derivatives of the \ac{NN}, we use automatic differentiation.
Therefore, evaluating the derivative of the loss functions requires a second-order automatic differentiation operation.

\subsection{Training}
\label{sec:train}

Our setup is an unusual case within machine learning as the dynamic training dataset is effectively of infinite size.
It lies within the regime of small data (the boundary dataset) with lots of physics (the \ac{DE} dataset)~\cite{Karniadakis2021}.

The two \acp{NN} for the real and imaginary parts of an integral family are trained independently.
Our implementation~\cite{repo,zenodo} uses the \texttt{PyTorch} framework~\cite{paszke2019pytorch}.
We use Adam optimisation~\cite{kingma2017adam}, generally with an initial learning rate of 0.001.
The \ac{NN} weights are initialised using the uniform Glorot scheme~\cite{pmlr-v9-glorot10a} with a gain of one.
Output biases are initialised as the means of the boundary values.

Analogously to mini-batch training, our training iterations are composed of small batches, taking a dynamic random sample of the inputs for each.
In other words, instead of having a fixed input sample, we have a different one for each iteration, which is computed on the fly by sampling the input variables according to some chosen phase-space distribution.
We find that the optimal batch size grows with the number of inputs and loop order, using a size between 64 and 256.
While an epoch no longer represents a complete sampling of the training dataset, it remains a useful concept for scheduling.
We thus group iterations into epochs, generally using 512--1024 iterations per epoch.
At the end of each epoch, we use the mean loss value of the epoch batches as the metric of training performance.
We also check if the epoch loss has plateaued and reduce the learning rate if it has.
We use the dynamic \texttt{ReduceLROnPlateau} scheduler for this, typically with a \textit{factor} of 0.1, \textit{threshold} of 0.01, a \textit{cooldown} of 0--5 epochs, and \textit{patience} of 3--6 epochs.\footnote{
    This scheduler keeps track of the minimum epoch loss, which we call \textit{best}.
    If, after the \textit{patience} number of epochs, the epoch loss does not drop below $\textit{best}\times(1+\textit{threshold})$, the learning rate is reduced by multiplying by the \textit{factor}.
    After the learning rate is reduced, the scheduler waits the \textit{cooldown} number of epochs before starting again.
}
Training is terminated when the learning rate reaches a minimum value of $10^{-8}$ and satisfies a patience of 3--10 epochs.

Since the inputs are dynamically randomly sampled, regularisation is built into the training scheme.
This means methods to avoid over-fitting like $L_2$ regularisation of the loss function or weight decay in the gradient descent algorithm are unnecessary.
Similarly, we use the training metric also as a validation metric rather than validating on a distinct dataset.\footnote{
    Since the training metric (epoch loss) is a mean squared difference, note that it cannot be directly compared between runs from different integral families or the same family with a different output distribution due to choices of kinematics or integral basis.
    For the latter, the differential error (\cref{eq:diff-err}) may be used.
}

\subsection{Inference uncertainty and testing error}
\label{sec:error}

Various sources of uncertainty have been studied in modelling~\cite{KIUREGHIAN2009105} and specifically deep learning~\cite{tagasovska2019singlemodel} and its application in high energy physics~\cite{Nachman:2019dol}.
We expect performance to be primarily limited by the flexibility of the optimisation procedure.

We adopt a simple method to obtain an uncertainty estimate on an inferred output: we train an ensemble~\cite{Ganaie_2022} of models, taking the mean value as the result and the standard error of the mean as its uncertainty~\cite{Nachman:2019dol,Badger:2020uow}.
Each model in the ensemble, which we refer to as a replica, is trained using a different random number seed for weight initialisation and training dataset generation.
This uncertainty estimate measures initialisation dependence and statistical uncertainty associated with the sampled phase space, although we expect the latter to be subdominant due to the dynamic sampling of $\ds_{\rm DE}$.
However, it is insensitive to any systematic bias in the result.
Its utility lies in the estimation of prediction precision, for instance to identify imprecise points within an output distribution.
In addition, the mean of the ensemble provides a more robust result than any single replica as it averages over the above stochastic effects.
Inspired by \incites{Badger:2020uow,Maitre:2021uaa,Maitre:2022xle}, we use ten replicas in the ensemble, finding it to provide a suitable distribution, albeit in an ad hoc manner.
More work will be necessary to establish the appropriate number of replicas.

During testing, we compare the testing dataset to model-inferred values on the testing inputs.
This may be statistically limited by the size of the testing dataset.
Only nonzero outputs are considered, since the globally zero coefficients may easily be identified from the boundary dataset so are known exactly.
For target $g$ and estimate $h$, we consider the following testing errors.
\begin{description}
    \item[Absolute difference]
        \begin{align}
            |g-h|
        \end{align}
        This can be useful to evaluate training performance since the loss function is also based on an absolute measure (mean squared error).

    \item[Magnitude of relative difference]
        \begin{align}
            |(g-h)/g|
        \end{align}
        This is a relative measure of the testing error, allowing comparison between different systems.

    \item[Logarithm of ratio]
        \begin{align}
            \log(h/g)
        \end{align}
        Similarly, there is also the natural logarithm of the ratio.
        We use its mean value as a relative measure of the testing accuracy.
    
    \item[Differential error]
        \begin{align}
            \label{eq:diff-err}
            \avesum{i=1}{\nikv-1} \left| \partial_{x_i} {h}_j^{(w)} - \sum_{k=0}^{\mathrm{min}(w,k_{\rm max})} \sum_{r=1}^{\nfn} A^{(k)}_{x_i,jr} \, {h}_r^{(w-k)} \right| \,,
        \end{align}
        for every $\eps$-order ($w=0,\ldots,w_{\text{max}}$) of every \ac{MI} ($j=1,\ldots,n_F$).
        Similarly to the \ac{DE} part of the loss function (\cref{eq:L-DE}), we can use \cref{eq:diff-err} to measure how well the model satisfies \cref{eq:DEs-finalform-f}.
        This is a useful quantity, for instance for high statistics validation, as it does not require the generation of the $g$ distribution.
\end{description}

\section{Examples}
\label{sec:Examples}

In this section, we present a number of applications of our method.
We warm up by resuming the one-loop massless box integral family introduced in \cref{sec:DE-Box}.
We then move on to two-loop order with three families of increasing complexity.
We train the \acp{NN} to approximate the \acp{MI} up to order $\eps^2$ at one loop, and $\eps^0$ at two loops. 
This is what is typically required for computing double virtual corrections.
We derive the analytic expression of the \acp{DE} using \texttt{LiteRed}~\cite{Lee:2012cn} and \texttt{FiniteFlow}~\cite{Peraro:2019svx}, as outlined in \cref{sec:DE-Box}.
The computation takes a few minutes on a laptop even for the most complicated case, and is thus not a bottleneck for our approach.
We spell out the propagators in \cref{app:families}, and provide the definitions of the \acp{MI} and the \acp{DE} they satisfy in the repository~\cite{repo,zenodo}.
All example datasets and trained models are available from \incite{models}.

For each example, we show representative learning curves for the training of the replicas of the real parts --- being the more difficult component --- in the ensemble.
Recall that since the training dataset is dynamically generated, the loss function is used for both the training and validation metrics.
The loss function is plotted by its mean value per epoch, with bands included to show the maximum and minimum value per epoch.
The replica training runs are sorted by the final loss value and, except for \cref{sec:topbox}, three runs are shown: the run with the lowest value, denoted ``Best''; the run with the closest to mean value, named ``Middle''; and the run with the highest value, as ``Worst''.
We also include the average training time for a single \ac{NN} as measured on a laptop.\footnote{All training timing tests were performed on an Nvidia GeForce RTX 3050 Ti Laptop GPU ($4\,\text{GB}$ RAM) using \texttt{PyTorch} with the \texttt{CUDA} backend.}

For examples with two-dimensional inputs, we plot the physical region after applying physical and spurious cuts, together with the points of the boundary dataset.

We perform testing on the trained models and report on the measured errors as defined in \cref{sec:error}.
We histogram the distributions of the testing errors, binning cumulatively over various subsets of the axes (real or imaginary part, function index, $\eps$ order) of the (nonzero) ensemble outputs.

In \cref{sec:GeneralComments}, we discuss a number of features which appear to be common to all examples, and gather a few tables detailing the hyperparameters we used and summarising the comparisons between our models and the testing datasets.

We collect a comprehensive set of plots detailing the training and testing statistics in \cref{app:plots}.
For the sake of conciseness, we give in the main text only selected figures which display noteworthy features.

\subsection{Massless box}
\label{sec:box}

\begin{figure}
    \centering
    \begin{subfigure}[c]{0.49\linewidth}
        \centering
        \includegraphics[width=\textwidth]{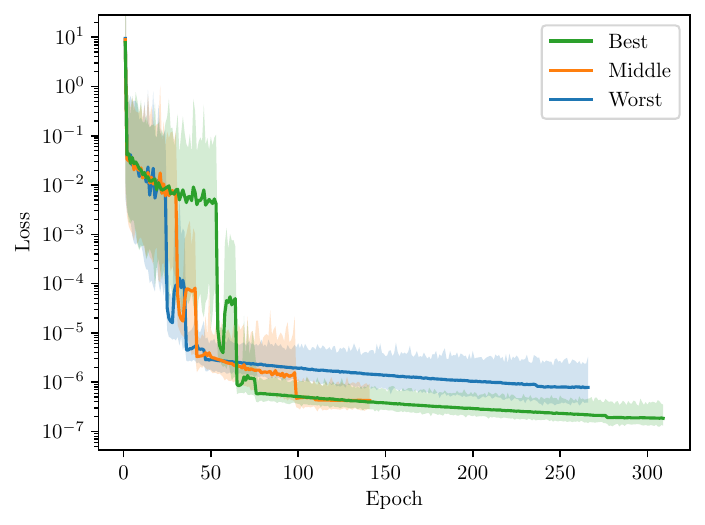}
        \caption{Learning curve.}
        \label{fig:training-box}
    \end{subfigure}
    \begin{subfigure}[c]{0.49\linewidth}
        \centering
        \includegraphics[width=\textwidth]{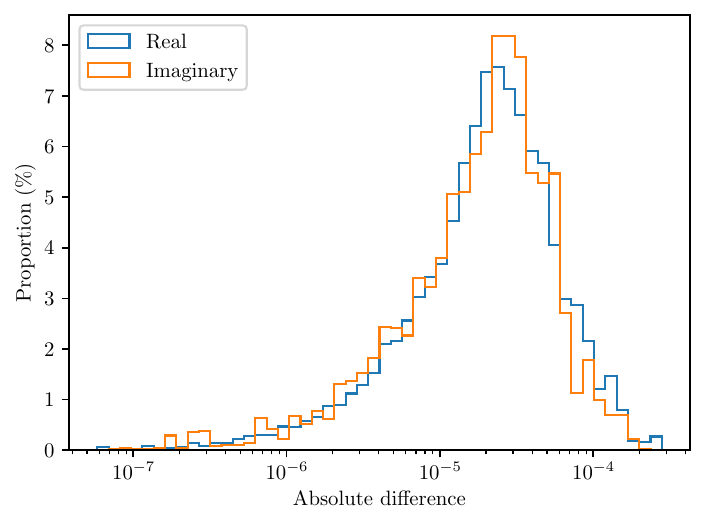}
        \caption{Absolute difference (real vs.\ imaginary).}
        \label{fig:abs-diff-part-box}
    \end{subfigure}

    \begin{subfigure}[c]{0.49\linewidth}
        \centering
        \includegraphics[width=\textwidth]{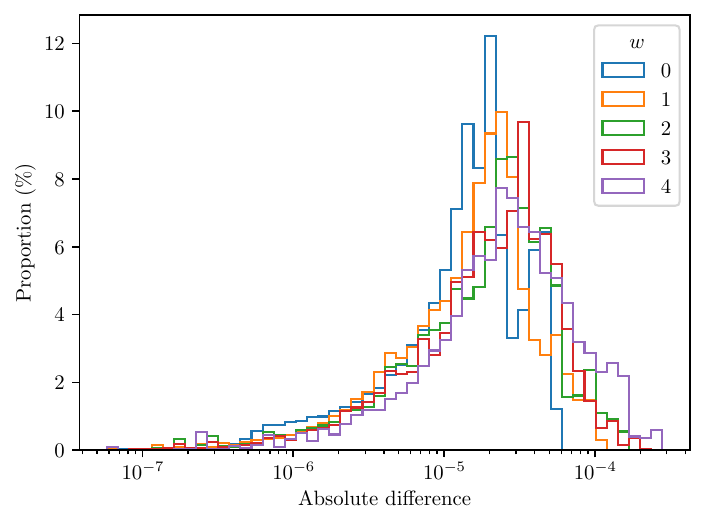}
        \caption{Absolute difference ($\eps$ orders).}
        \label{fig:abs-diff-eps-box}
    \end{subfigure}
    \begin{subfigure}[c]{0.49\linewidth}
        \centering
        \includegraphics[width=\textwidth]{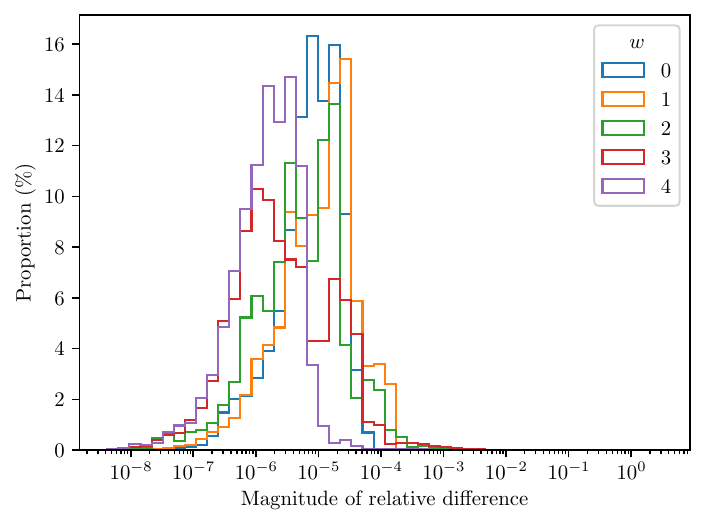}
        \caption{Magnitude of relative difference ($\eps$ orders).}
        \label{fig:rel-err-dist-box}
    \end{subfigure}

    \begin{subfigure}[c]{0.49\linewidth}
        \centering
        \includegraphics[width=\textwidth]{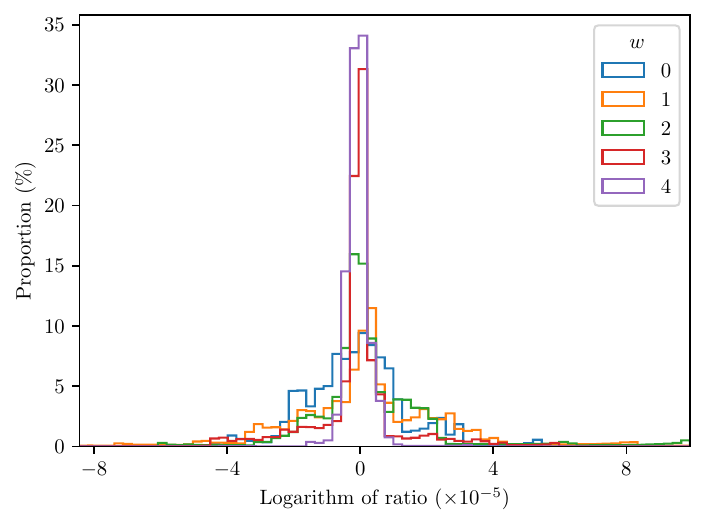}
        \caption{Logarithm of ratio ($\eps$ orders).}
        \label{fig:ratio-box}
    \end{subfigure}
    \begin{subfigure}[c]{0.49\linewidth}
        \centering
        \includegraphics[width=\textwidth]{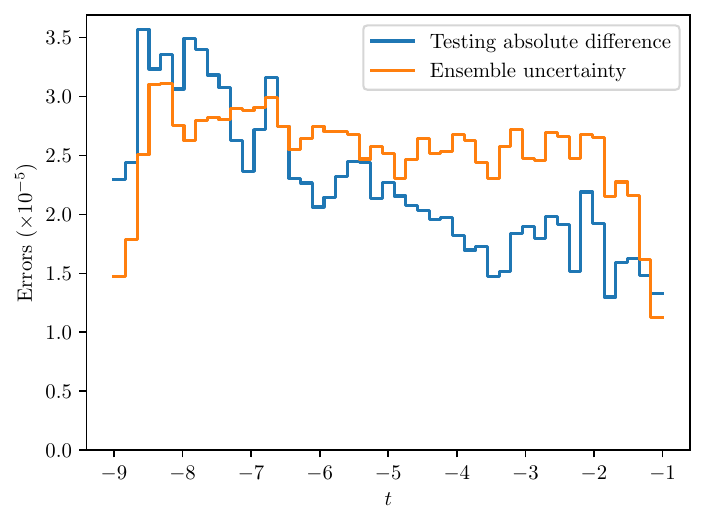}
        \caption{Errors binned over the phase space.}
        \label{fig:err-ps-box}
    \end{subfigure}

    \caption{
        Training, testing, and inference statistics for the massless box.
        In \cref{fig:err-ps-box}, $t$ is the kinematic invariant defined in \cref{sec:DE-Box}.
    }
    \label{fig:box-stats}
\end{figure}

We refer to \cref{sec:DE-Box} for all definitions of the massless box.
We recall that this family has $\nfn=3$ \acp{MI}, which depend on $\nikv=2$ kinematic variables, $\vec{v}=\{s,t\}$.
Since we do not want to rely on the canonical form of the \acp{DE}, we use the \acp{MI} defined in \cref{eq:box-noncanMI}.
We further multiply them by suitable factors of $s$ to cancel their scaling dimensions.
The connection matrices are linear in $\eps$ (hence $k_{\rm max}=1$).
The $\eps$ expansion of the \acp{MI} in \cref{eq:Fexp} starts from $w^* = -2$, and we truncate it at $\eps^{2}$, thus considering $n_{\eps} = 5$ orders ($w_{\rm max} = 4$).

We work in the $s$ channel, where $s>-t>0$.
Both the physical singularities ($s=0$ and $t=0$) and the spurious singularity ($s+t = 0$) are at the boundary of the $s$ channel.
No cuts are thus needed to remove the spurious singularities from the \ac{DE} dataset.
The cut on the physical singularities is $10 \%$ of $s$, which we fix to $c_s = 10$.
The \acp{NN} thus have only one input, $\vec{x} = ( c_s \, t/s )$, and $\nfn \, \neps = 15$ outputs.
We find that 3 hidden layers of 32 nodes each are sufficient to fit the solution.
The relevant values of the hyperparameters are summarised in \cref{tab:hyperparameters}.

In order to obtain the required boundary values and test our model, we solve the \acp{DE} analytically in terms of \acp{MPL} (see e.g.\ \cref{eq:box-sol}) in the Euclidean region, and continue the solution to the $s$-channel by adding infinitesimal positive imaginary parts to $s$ and $t$.
We evaluate the \acp{MPL} using \texttt{handyG}~\cite{Naterop:2019xaf}.
The boundary dataset comprises 2 points taken on the edge points of the cut $s$ channel.

In \cref{fig:training-box} we show the learning curve of the ensemble, namely the value of the loss function for the real parts evaluated on the training dataset (see \cref{sec:train}).
Training of a single \ac{NN} takes 16 minutes on average.
In \cref{fig:abs-diff-part-box,fig:abs-diff-eps-box,fig:rel-err-dist-box,fig:ratio-box} we detail in various ways the comparison between our ensemble and the analytic results on a testing dataset of $10^5$ points.
\Cref{fig:abs-diff-part-box} shows a histogram of the absolute difference, distinguishing real and imaginary parts.
Despite the real parts being substantially more complicated, the \acp{NN} perform similarly well for them as for the imaginary parts.
A similar observation can be made for the different orders in $\eps$ of the solution, as seen in \cref{fig:abs-diff-eps-box}.
The cumulative mean absolute difference is $2.9\times10^{-5}$.
In \cref{fig:rel-err-dist-box,fig:ratio-box} we show the magnitude of the relative difference and the logarithm of the ratio, distinguishing the orders in $\eps$.
The analytic complexity grows with the order in $\eps$, yet the \acp{NN} perform comparably well at all orders.
The logarithmic ratio histograms are narrow and closely zero-centred, albeit with a small number of outliers.
The logarithmic ratio has a cumulative mean value of $3.9\times10^{-7}$, while the overall mean magnitude of relative difference is $2.2\times10^{-5}$.
Finally, in \cref{fig:err-ps-box} we compare the ensemble uncertainty against the testing absolute difference.
In this case, the ensemble uncertainty appears to be a good estimate of the uncertainty of the fit, as it is compatible with the testing error throughout the phase-space region under consideration.
We recall however that the ensemble uncertainty is expected to catch only the part of the uncertainty due to the random initialisation of the \acp{NN}.
We discuss the problem of estimating the uncertainty within our method in \cref{sec:GeneralComments}.

\subsection{Two-loop four-point planar with an external mass}
\label{sec:1m-doublebox}

\begin{figure}
        \centering
        \begin{tikzpicture}
        \node at (0,0) {\includegraphics[scale=0.7]{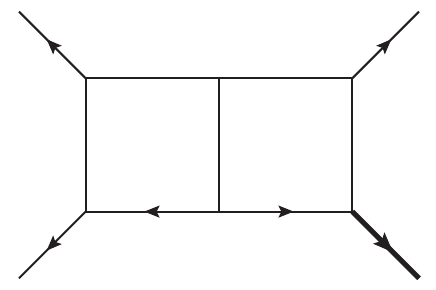}};
        \node at (-2.15,-1) {$p_1$};
        \node at (-2.15,1) {$p_2$};
        \node at (2.25,1) {$p_3$};
        \node at (2.25,-1) {$p_4$};
        \node at (-0.7,-1.1) {$k_1$};
        \node at (0.85,-1.1) {$k_2$};
    \end{tikzpicture}
    \caption{
        Graph representing the propagator structure of the one-mass double box integral family.
        The arrows denote the directions of the momenta.
        Bold lines are massive.
    }
    \label{fig:1m-doublebox}
\end{figure}
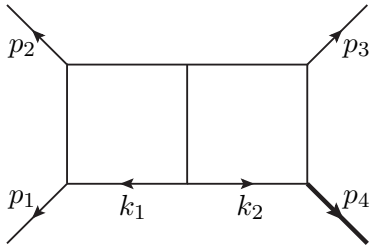
  
\begin{figure}
\centering
      \begin{subfigure}[c]{0.49\linewidth}
          \centering
          \includegraphics[width=0.75\textwidth]{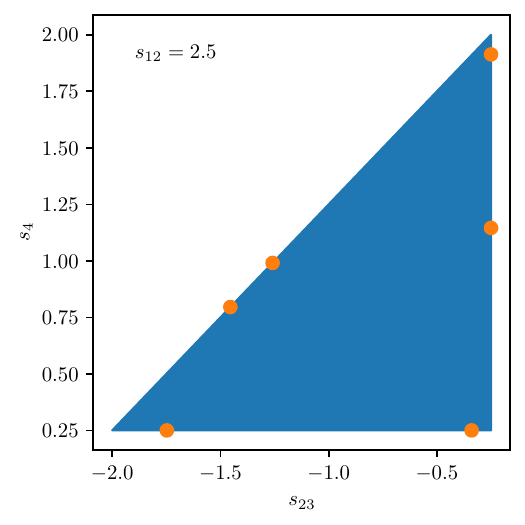}
          \caption{
              Scattering region with boundary points.
          }
          \label{fig:region-plot-t331ZZZM}
    \end{subfigure}
      \begin{subfigure}[c]{0.49\linewidth}
        \centering
        \includegraphics[width=\textwidth]{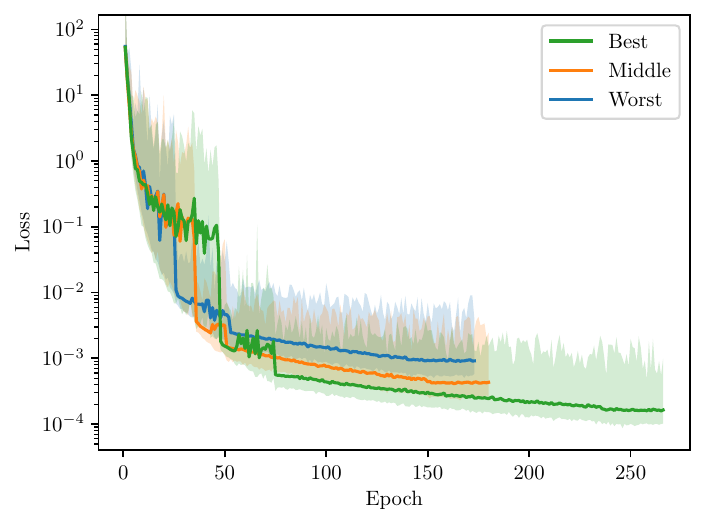}
        \caption{Learning curve.}
        \label{fig:training-t331ZZZM}
    \end{subfigure}
    \begin{subfigure}[c]{0.49\linewidth}
        \centering
        \includegraphics[width=\textwidth]{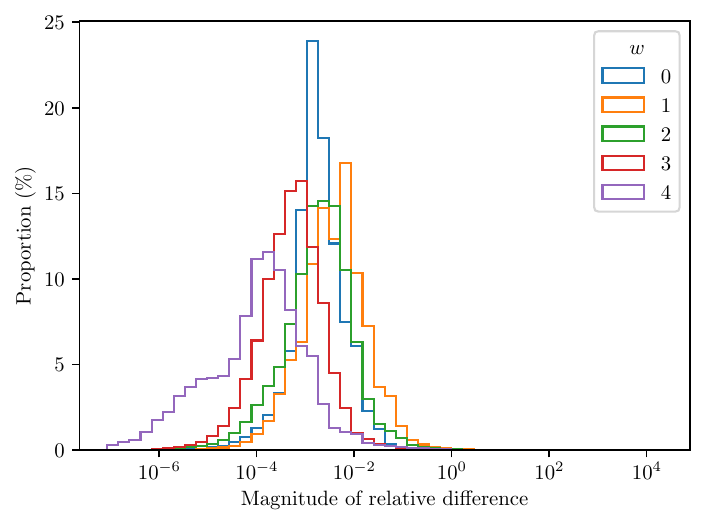}
        \caption{Magnitude of relative difference ($\eps$ orders).}
        \label{fig:rel-diff-t331ZZZM}
    \end{subfigure}
    \begin{subfigure}[c]{0.49\linewidth}
        \centering
        \includegraphics[width=\textwidth]{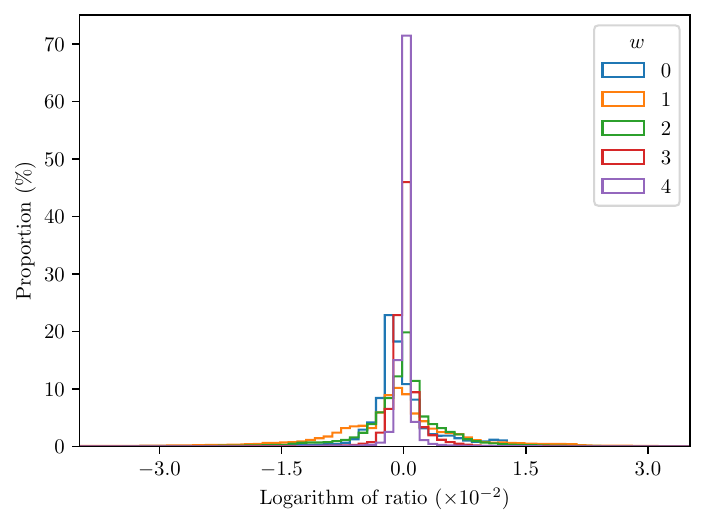}
        \caption{Logarithm of ratio ($\eps$ orders).}
        \label{fig:ratio-t331ZZZM}
    \end{subfigure}
   
    \caption{
        Training and testing statistics for the one-mass double box.
    }
    \label{fig:dblbox-stats-body}
\end{figure}

We now consider the family of Feynman integrals represented by the graph in \cref{fig:1m-doublebox}.
We dub it \emph{one-mass double box}.
These integrals have been computed analytically in \incites{Gehrmann:2000zt,Gehrmann:2002zr,Duhr:2012fh,Badger:2023xtl,Fadin:2023phc,Gehrmann:2023etk}.
We use the results of \incite{Badger:2023xtl} to construct the boundary dataset and test the model.

There are $\nfn=18$ \acp{MI}, which depend on $\nikv=3$ kinematic variables.
We choose $\vec{v} = \{ s_{12}, s_{23}, s_4 \}$, with $s_{ij} = (p_i+p_j)^2$ and $s_4 = p_4^2$ ($p_i^2 = 0$ for $i=1,2,3$).
We use the \ac{MI} basis of \incite{Badger:2023xtl}, but remove the normalisation factors which make it canonical.
The connection matrices are then linear in $\eps$ ($k_{\rm max}=1$).
The $\eps$ expansion of the \acp{MI} in \cref{eq:Fexp} starts from $w^* = -4$, and we truncate it at $\eps^{0}$, thus considering $n_{\eps} = 5$ orders ($w_{\rm max} = 4$).
In this case we do not observe any benefit from cancelling the scaling dimensions of the \acp{MI}, and thus refrain from doing so.

We consider the $s_{12}$ channel:
\begin{align}
    s_{12} > s_4 - s_{23} \, \land \, s_{23}<0 \, \land \, s_4 > 0 \,.
\end{align}
The physical singularities are at the boundary of this region,
\begin{align}
    s_{12} = 0\,, \qquad \quad s_{23} = 0 \,, \qquad \quad s_4 = 0 \,, \qquad \quad s_{12}+s_{23}-s_4 = 0 \,.
\end{align}
We impose a cut on them equal to $10\%$ of $s_{12}$, which we fix to $c_{s_{12}} = 2.5$.
The spurious singularities are
\begin{align}
    s_{12} - s_4 = 0\,, \qquad \quad s_{23} - s_4 = 0 \,, \qquad \quad s_{12}+s_{23} = 0 \,.
\end{align}
They lie outside of the $s_{12}$ channel and thus do not need to be cut.
The boundary dataset comprises 6 points distributed on the boundary of the cut $s_{12}$ channel, as shown in \cref{fig:region-plot-t331ZZZM}.

The step-up in difficulty of the one-mass double box is considerable compared to the one-loop massless box.
The model input is now two-dimensional, while the output has a 6 times greater size and describes more complicated functions.
However, our model still trains well as shown by the learning curve in \cref{fig:training-t331ZZZM}.
Training of a single \ac{NN} takes around 53 minutes.
The observations made for the box in \cref{sec:box} hold true here too: real and imaginary parts, as well as the different orders in $\eps$, are learnt by our model equally well.
For example, we compare our model against the analytic solutions order by order in $\eps$ at $10^5$ points in \cref{fig:rel-diff-t331ZZZM,fig:ratio-t331ZZZM}, showing the magnitude of the relative difference and the logarithm of the ratio, respectively.
Our model achieves a mean magnitude of relative difference of $1.1\times10^{-2}$, with a mean logarithm of ratio of $-2.8\times10^{-4}$.
See \cref{app:t331ZZZM} for further plots.

\subsection{Two-loop four-point non-planar with a closed massive loop}
\label{sec:heavycrossedbox}

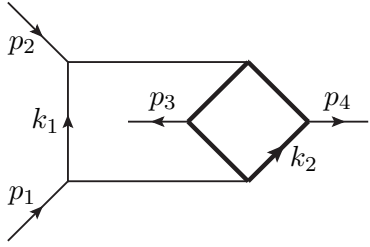
\begin{figure}
    \centering
    \begin{tikzpicture}
        \node at (0,0) {\includegraphics[scale=0.7]{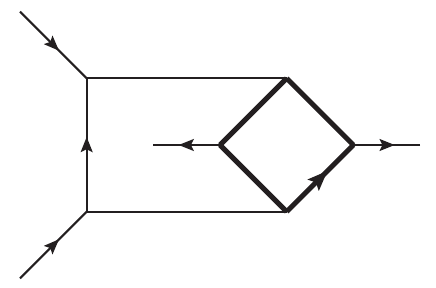}};
        \node at (-2.15,-1) {$p_1$};
        \node at (-2.15,1) {$p_2$};
        \node at (-0.3,0.25) {$p_3$};
        \node at (2,0.25) {$p_4$};
        \node at (-1.85,0) {$k_1$};
        \node at (1.55,-0.5) {$k_2$};
    \end{tikzpicture}
    \caption{
        Graph representing the propagator structure of the heavy crossed box integral family.
        The arrows denote the directions of the momenta.
        Bold lines are massive.
    }
    \label{fig:heavycrossedbox}
\end{figure}

\begin{figure}
    \centering
      \begin{subfigure}[c]{0.49\linewidth}
        \centering
        \includegraphics[width=0.75\textwidth]{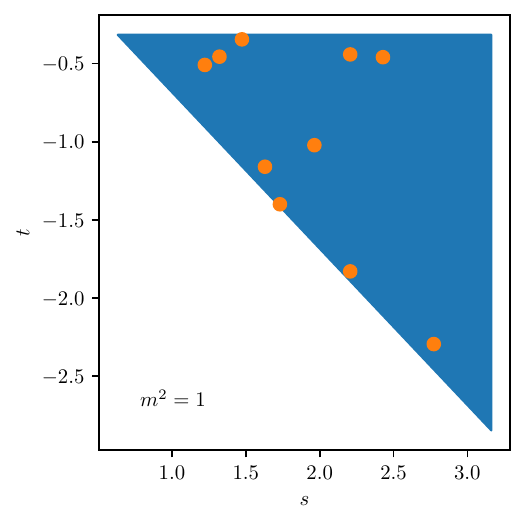}
        \caption{
            Scattering region with boundary points.
        }
        \label{fig:region-plot-hcb}
      \end{subfigure}
      \begin{subfigure}[c]{0.49\linewidth}
        \centering
        \includegraphics[width=\textwidth]{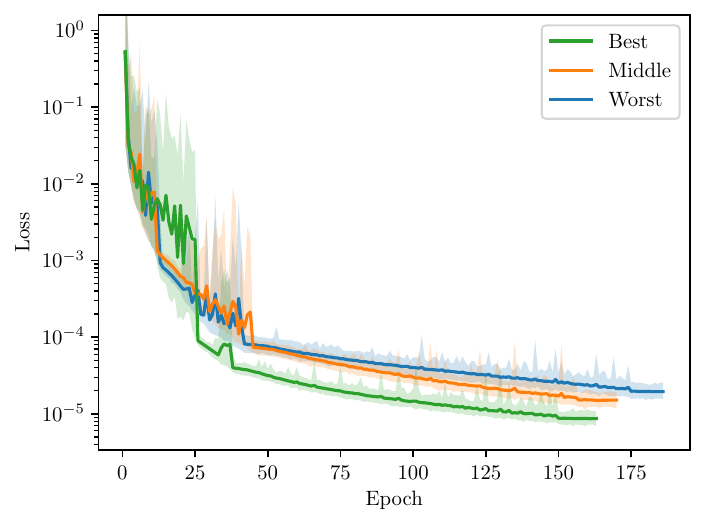}
        \caption{Learning curve.}
        \label{fig:training-heavycrossbox}
    \end{subfigure}
    \begin{subfigure}[c]{0.49\linewidth}
        \centering
        \includegraphics[width=\textwidth]{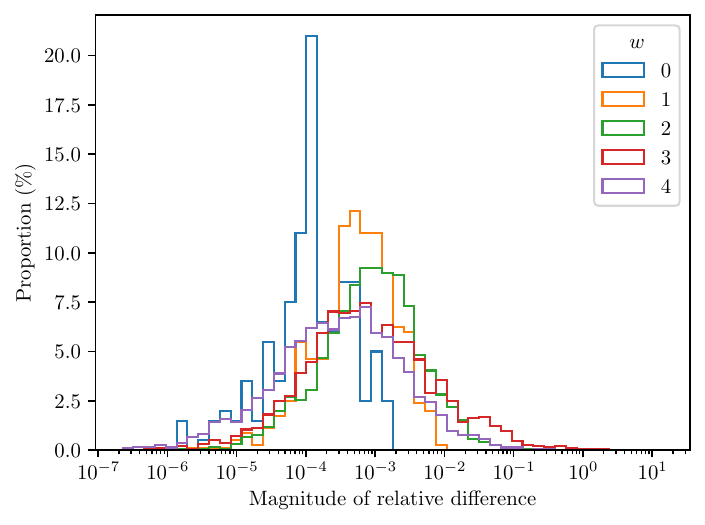}
        \caption{Magnitude of relative difference ($\eps$ orders).}
        \label{fig:rel-diff-heavycrossbox}
    \end{subfigure}
    \begin{subfigure}[c]{0.49\linewidth}
        \centering
        \includegraphics[width=\textwidth]{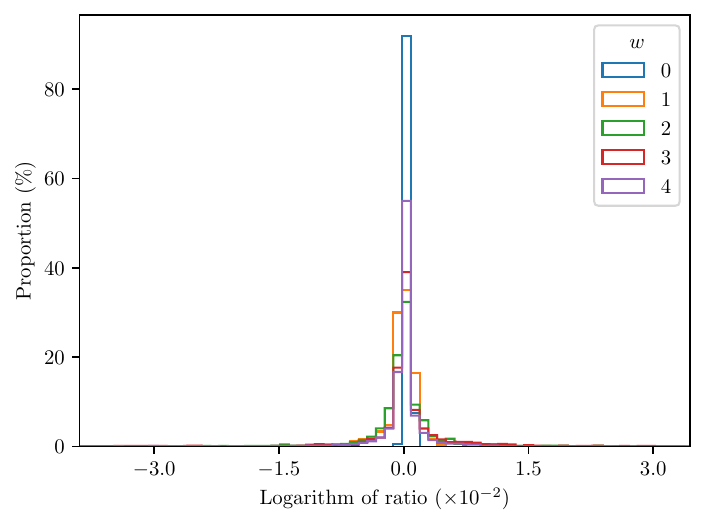}
        \caption{Logarithm of ratio ($\eps$ orders).}
        \label{fig:ratio-heavycrossbox}
    \end{subfigure}

    \caption{
        Training and testing statistics for the heavy crossed box.
    }
    \label{fig:hcb-stats-body}
\end{figure}

In this section, we study the family of non-planar two-loop four-point integrals represented by the graph in \cref{fig:heavycrossedbox}.
We dub this family \emph{heavy crossed box}.
With respect to the two-loop four-point integrals discussed in the previous sections, the heavy crossed boxes are non-planar and have a closed loop of massive propagators.
The jump in complexity, from the analytic point of view, is substantial. 
This is in fact the first example we present where the solution cannot be expressed in terms of \acp{MPL}.
While an analytic expression in terms of elliptic \acp{MPL}~\cite{Broedel:2019hyg} has been obtained in \incite{vonManteuffel:2017hms} for a subsector (i.e., for a subset of integrals where some of the propagator powers are set to zero), the analytic expression of \emph{all} \acp{MI} and a system of \acp{DE} in canonical form are still unknown.
A full computation was completed only very recently~\cite{Becchetti:2023wev,Becchetti:2023yat} by solving the \acp{DE} \emph{numerically} with the method of generalised power series expansions~\cite{Moriello:2019yhu} implemented in \texttt{DiffExp}~\cite{Hidding:2020ytt}.
To evaluate these integrals numerically in order to create the boundary dataset and test our model we use \texttt{AMFlow}~\cite{Liu:2017jxz,Liu:2021wks,Liu:2022chg} with a target precision of 16 digits.
While the latter can evaluate these integrals easily, with our method we aim at achieving an evaluation time that is several orders of magnitude shorter.

There are $\nfn=36$ \acp{MI}, which depend on $\nikv=3$ kinematic variables.
We choose $\vec{v} = \{ s, t, m^2 \}$, with $s = (p_1+p_2)^2$ and $t = (p_1-p_3)^2$, while $m^2$ is the mass of the internal propagators.\footnote{Note that here $p_1$ and $p_2$ are taken as ingoing, so momentum conservation reads $p_1+p_2 = p_3 + p_4$.}
We use a modification of the \ac{MI} basis of \incite{Becchetti:2023wev}.
We remove all normalisation factors which contain square roots, so that the connection matrices are rational functions of the kinematics, and multiply the \acp{MI} by suitable factors of $m^2$ to cancel their scaling dimensions.\footnote{The connection matrices in \incite{Becchetti:2023wev} have a factor of $(1+2 \eps)$ in the denominator. 
We remove the latter and make the connection matrices polynomial in $\eps$ by multiplying the $32^{\rm nd}$ \ac{MI} by $(1+2 \eps)$.}
The connection matrices are polynomial in $\eps$ up to order $k_{\rm max}=2$, in contrast with the previous two examples where they were linear.
The $\eps$ expansion of the \acp{MI} in \cref{eq:Fexp} starts from $w^* = -4$, and we truncate it at $\eps^{0}$, thus considering $n_{\eps} = 5$ orders ($w_{\rm max} = 4$).

We consider the $s_{12}$ channel,
\begin{align}
    s>-t>0 \, \land \, m^2 > 0.
\end{align}
We fix $m^2$ to $c_{m^2} = 1$.
The physical singularities are where any of the following $4$ factors vanish,
\begin{align}
\bigl\{ m^2 \,, \ s \,, \ t \,, \ s+t \bigr\} \,,
\end{align}
while the spurious singularities are associated with the following $9$ factors,
\begin{align}
\begin{aligned}
\bigl\{ & s - 4 m^2 \,, \  s + 4 m^2  \,, \ s + 16 m^2 \,, \ s + t + 4 m^2 \,, \  t - 4 m^2 \,, \\ 
& t - 8 m^2 \,, \ s (s + t) - 4 m^2 t  \,, \ t (s + t) - 4 m^2 s \,, \ s t - 4 m^2 (s + t) \bigr\} \,.
\end{aligned}
\end{align}
The first spurious singularity, $s = 4 m^2$, crosses the physical region. 
We prefer to postpone the treatment of spurious singularities to the next example, and focus here on the additional complexity due to the non-polylogarithmic special functions involved in the solution.
This allows for a more fair comparison against the results of the previous section.
For this reason we restrict the range of $s$ to $s < \sqrt{10}$, this way excluding the spurious singularity from the domain.
We take 10 boundary points, randomly sampled from the scattering region as shown in \cref{fig:region-plot-hcb}.

As for the previous examples, we plot the decay of the loss function as our learning curve in \cref{fig:training-heavycrossbox}. This time the training of a single \ac{NN} takes around 75 minutes.
The logarithmic ratio has a cumulative mean of absolute value $4.5 \times 10^{-4}$ and the mean magnitude of the relative difference amounts to $7.3 \times 10^{-3}$.
In \cref{fig:rel-diff-heavycrossbox} and \cref{fig:ratio-heavycrossbox} we respectively show the magnitude of the relative difference and the logarithm of the ratio, in the various $\eps$ orders. Again we note how the performance does not depend on the analytic complexity, which increases with the order in the $\eps$ expansion.
The fact that the order $\eps^0$ appears to perform better than the higher orders in \cref{fig:rel-diff-heavycrossbox} is misleading. 
One should in fact note that, conversely, the order $\eps^4$ is doing better than the simpler order $\eps^3$.
In general, the performance of the model for the single \ac{MI} coefficient is uncorrelated to its analytic complexity.
See \cref{app:heavycrossbox} for further plots.

\subsection{Two-loop four-point planar for top-pair production with a closed top loop}
\label{sec:topbox}

\begin{figure}
    \centering
    \begin{tikzpicture}
        \node at (0,0) {\includegraphics[scale=0.7]{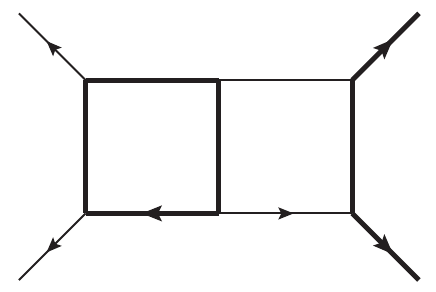}};
        \node at (-2.15,-1) {$p_1$};
        \node at (-2.15,1) {$p_2$};
        \node at (2.25,1) {$p_3$};
        \node at (2.25,-1) {$p_4$};
        \node at (-0.7,-1.1) {$k_1$};
        \node at (0.85,-1.1) {$k_2$};
    \end{tikzpicture}
    \caption{
        Graph representing the propagator structure of the top double box integral family.
        The arrows denote the directions of the momenta.
        Bold lines are massive.
    }
    \label{fig:topbox}
\end{figure}
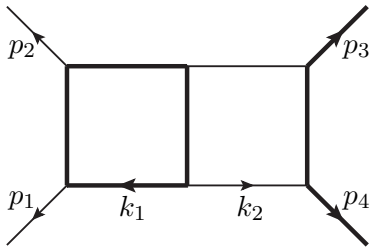

\begin{figure}
    \centering
     \begin{subfigure}[c]{0.49\linewidth}
        \centering
        \includegraphics[width=0.75\textwidth]{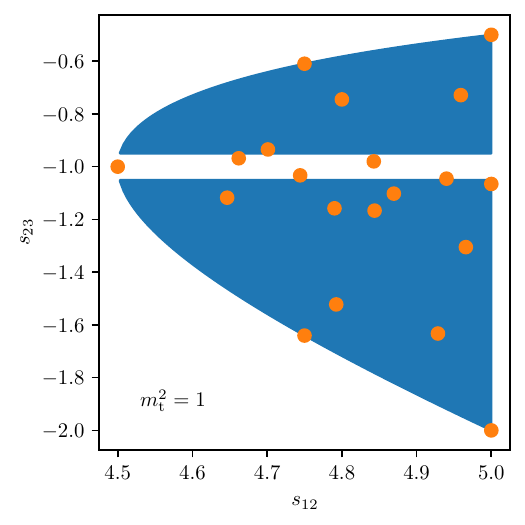}
        \caption{
            Scattering region with boundary points.
        }
        \label{fig:region-plot-tb}
    \end{subfigure}
      \begin{subfigure}[c]{0.49\linewidth}
        \centering
        \includegraphics[width=\textwidth]{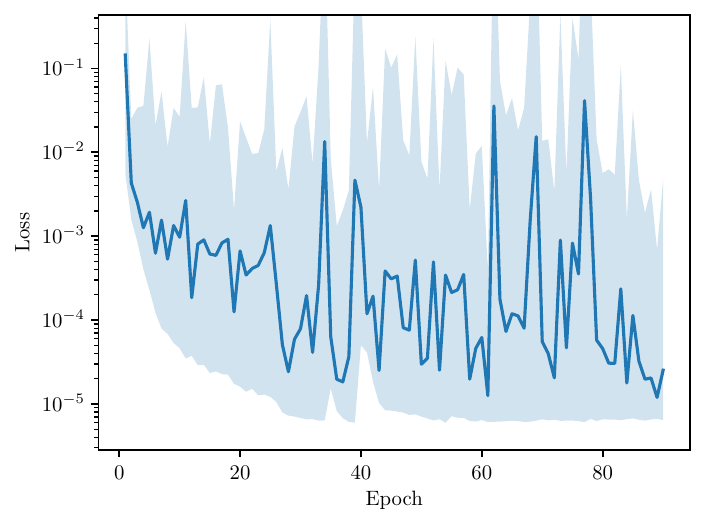}
        \caption{Learning curve.}
        \label{fig:training-topbox}
    \end{subfigure}
    \begin{subfigure}[c]{0.49\linewidth}
        \centering
        \includegraphics[width=\textwidth]{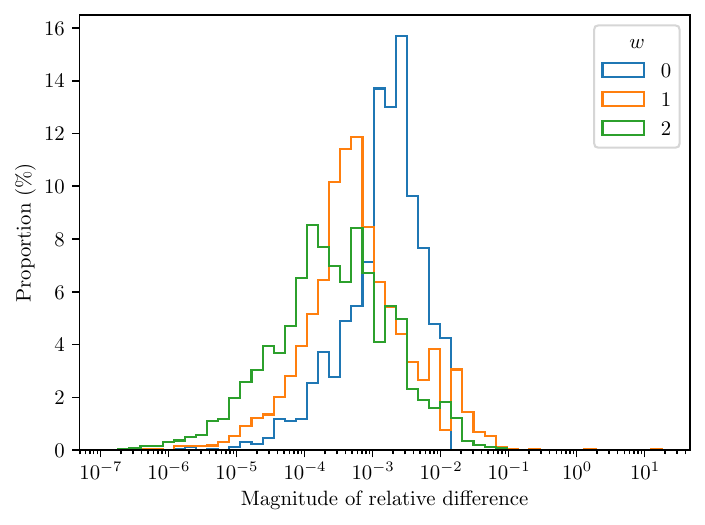}
        \caption{Magnitude of relative difference ($\eps$ orders).}
        \label{fig:rel-diff-topbox}
    \end{subfigure}
    \begin{subfigure}[c]{0.49\linewidth}
        \centering
        \includegraphics[width=\textwidth]{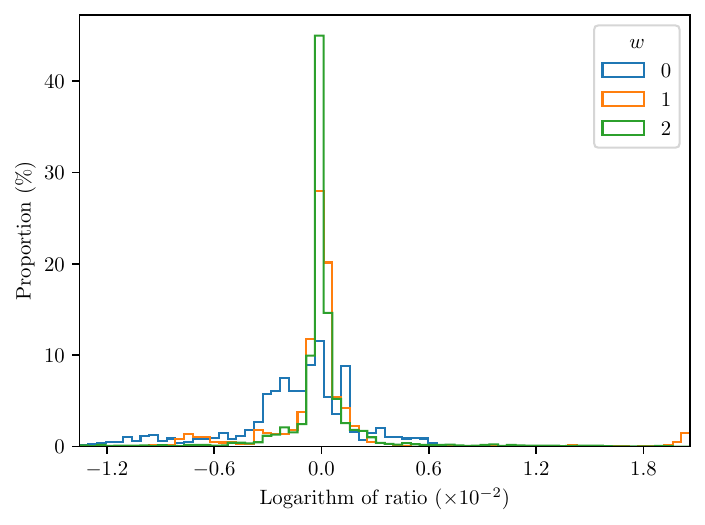}
        \caption{Logarithm of ratio ($\eps$ orders).}
        \label{fig:ratio-topbox}
    \end{subfigure}

    \caption{
        Training and testing statistics for the top double box.
    }
    \label{fig:tb-stats-body}
\end{figure}

Our final and most complicated example is the family of integrals associated with the graph in \cref{fig:topbox}, which we dub \emph{top double box}.
The top double box integrals were first computed by solving numerically the \acp{DE}~\cite{Czakon:2013goa,Barnreuther:2013qvf}.
The analytic study of \incites{Adams:2018kez,Adams:2018bsn} later revealed an extreme degree of analytic complexity.
The presence of three inequivalent elliptic curves puts this integral family at the cutting edge of analytic computations.
Indeed, a canonical form of the \acp{DE} has become available recently only for a subsector~\cite{Muller:2022gec}.
In \incites{Adams:2018kez,Adams:2018bsn}, instead, the authors obtained connection matrices linear in $\eps$, and with a block triangular structure which allows for a solution in terms of Chen iterated integrals~\cite{Chen:1977oja}.
Nonetheless, the numerical evaluation of these functions remains challenging (see e.g.~\cite{Badger:2021owl}).
As in the previous example, therefore, we rely on \texttt{AMFlow} for evaluating the \acp{MI} numerically, using a target precision of 10 digits.

There are $44$ \acp{MI}, which depend on $\nikv=3$ kinematic variables.
We choose $\vec{v} = \{ s_{12}, s_{23}, m_{\rm t}^2 \}$, with $s_{ij} = (p_i+p_j)^2$ and $p_3^2 = p_4^2 = m_{\rm t}^2$ ($p_1^2 = p_2^2 = 0$).
We constructed a \ac{MI} basis such that the dependence on $\eps$ is factorised from the dependence on the kinematics in the denominators of the connection matrices~\cite{Smirnov:2020quc,Usovitsch:2020jrk},\footnote{We thank Tiziano Peraro for constructing this basis with his own algorithm.} and normalised some of the \acp{MI} by factors of $\eps$ to make the connection matrices finite at $\eps=0$.
The resulting connection matrices are rational functions of $\eps$.
We Taylor expand them around $\eps = 0$.
Note that the internal mass regulates the infrared divergences.
As a result, the $\eps$ expansion of the \acp{MI} in \cref{eq:Fexp} starts from $w^* = -2$ rather than $-4$ as in the other two-loop examples.
By truncating the expansion at $\eps^{0}$ we then consider $n_{\eps} = 3$ orders ($w_{\rm max} = 2$).
For this we only need the connection matrices up to $\eps^2$.
The higher orders are neglected.
From the boundary dataset we can also determine that, up to $\eps^0$, $11$ of the $44$ \acp{MI} vanish identically.
We thus eliminate them from the system, leaving $\nfn=33$ nonzero \acp{MI}.

We consider the $s_{12}$ channel,\footnote{The quadratic constraint comes from a Gram determinant of three momenta.}
\begin{align}
s_{12} > 4 m_{\rm t}^2 > 0 \, \land \,  s_{23} (s_{12} + s_{23}) - 2 m_{\rm t}^2 s_{23} + m_{\rm t}^4 < 0 \,.
\end{align}
We fix $m_{\rm t}^2$ to $c_{m_{\text{t}}^2} = 1$, and vary $s_{12}$ in $(0, 5)$.
The physical singularities are associated with the following $4$ factors,
\begin{align}
\bigl\{ m_{\rm t}^2 \,, \ s_{12} \,, \ s_{23} \,, \  s_{23} (s_{12} + s_{23}) - 2 m_{\rm t}^2 s_{23} + m_{\rm t}^4 \bigr\} \,,
\end{align}
and there is a spurious singularity where any of the following $8$ factors vanish,
\begin{align}
\begin{aligned}
\bigl\{ & m_{\rm t}^2 + s_{23} \,, \ s_{12} + 4 m_{\rm t}^2 \,, \  s_{23} - m_{\rm t}^2 \,, \ s_{12} - 4 m_{\rm t}^2 \,, \ s_{23} - 9 m_{\rm t}^2 \,, \ s_{12} + s_{23} - m_{\rm t}^2 \,, \\
& s_{12} s_{23} - 9 m_{\rm t}^2 s_{12} - 4 m_{\rm t}^2 s_{23} + 4 m_{\rm t}^4 \,, \ 64 m_{\rm t}^6 - 9 m_{\rm t}^4 s_{12} + 10 m_{\rm t}^2 s_{12} s_{23} - s_{12} s_{23}^2 \bigr\}\,.
\end{aligned}
\end{align}
The first spurious singularity, $s_{23} + m_{\rm t}^2 = 0$, crosses the $s_{12}$ channel.
Therefore, in addition to the cut on the physical singularities at the boundary of the $s_{12}$ channel, we also need a cut to exclude the spurious singularity from the \ac{DE} dataset. 
We choose $10 \%$ of the largest value of $s_{12}$ for the former, and $1\%$ for the latter.
The shape of the $s_{12}$ channel and the spurious cut crossing it are shown in \cref{fig:region-plot-tb}.
We stress that this is a novel feature, absent in the previous examples.
We take 20 boundary points, with 5 manually placed on the boundaries and the remaining 15 randomly sampled from the interior of the physical~region.

While for the previous examples we found that three hidden layers were sufficient, we see a performance improvement by moving to four layers for the top double box.
We plot the learning curve in \cref{fig:training-topbox}, where the difficulty in training at this level of analytic complexity is revealed by the instability of the loss function.
Indeed, in order to make the figure clearer, in this case we plot only the ``best'' learning curve, namely the one ending with the lowest loss value.
The other curves are in this case qualitatively the same. 
We observe that this is a result of a small fraction of training points being pathological, which was not regulated by increasing the cuts.
The learning curve ends in a region which is not visually a plateau due to a setting that terminates training a fixed number of epochs after the learning rate has reached the minimum value of $10^{-8}$ (see \cref{sec:train}). 
This time the training of a single \ac{NN} takes around 32 minutes.
In testing, the performance of the basis functions varied by up to an order of magnitude as measured by the mean absolute difference per basis function.
However, the average fit is good, with a cumulative mean of the logarithmic ratio of $1.4 \times 10^{-3}$ and a mean magnitude of the relative difference of $3.9\times10^{-3}$.
\Cref{fig:rel-diff-topbox,fig:ratio-topbox} respectively show the distributions of the magnitude of the relative difference and the logarithm of the ratio, in the successive $\eps$ orders.
See \cref{app:topbox} for further plots.

For this example, we found $\approx 50\%$ ($\approx 40\%$) of the replica trainings for the real (imaginary) part to fail, having a differential error in validation up to a few times larger than the others.
Given the exploratory nature of this study, we decided to train a larger number of replicas, and keep in the final ensemble the ten which perform the best according to the differential error.
Further study is required to understand the onset of this phenomenon, and mitigate it.
For example, improvements in the optimisation algorithm have been shown to reduce the fraction of discarded replicas from $30\%$ to $1\%$ in Parton-Distribution-Function fits~\cite{NNPDF:2021njg}.
Furthermore, more sophisticated algorithms for selecting replicas to boost the performance of the ensemble are available in the machine-learning literature (see e.g.~\incite{ZHOU2002239}), and it would be interesting to test them in our approach as well.

\subsection{General comments}
\label{sec:GeneralComments}

Before discussing the general features we observed in the previous examples, we collect here a number of common tables to facilitate the comparison. 
In \cref{tab:hyperparameters} we summarise the hyperparameters we used, 
\cref{tab:training} gives some statistics on the training,
and \cref{tab:testing} describes the comparison between our models and the testing dataset.
We recall that \cref{app:plots} collects a comprehensive set of plots for all two-loop examples, mirroring \cref{fig:box-stats} for the one-loop massless box.
We emphasise that the last two examples (the heavy crossed box and the top double box) have lower statistics than the first two (the massless box and the one-mass double box), because we had to rely on \texttt{AMFlow} for the numerical evaluation of the integrals in the former, as opposed to the analytic results available for the latter.

\begin{table}
    \centering
    \begin{tabular}{|c|c c c c|}
        Integral family & box & one-mass double box & heavy crossed box & top double box \\
        Inputs & 1 & 2 & 2 & 2 \\
        Hidden layers & $3\times32$ & $3\times256$ & $3\times256$ & $4\times128$ \\
        Outputs & 15 & 90 & 180 & 99 \\
        Learning rate & $10^{-2}$ & $10^{-3}$ & $10^{-3}$ & $10^{-3}$ \\
        Batch size & 64 & 256 & 256 & 256 \\
        Boundary points & 2 & 6 & 10 & 20 \\
        $c_{\nikv}$ & $s=10$ & $s_{12}=2.5$ & $m^2=1$ & $m_\text{t}^2=1$ \\
        Scale bound & --- & --- & $s\le\sqrt{10}$ & $s_{12}\le5$ \\
        Physical cut (\%) & 10 & 10 & 10 & 10 \\
        Spurious cut (\%) & 0 & 0 & 0 & 1 \\
    \end{tabular}
    \caption{
        Summary of hyperparameters for all examples, showing
            \ac{NN} architecture (inputs, hidden layers depth $\times$ width, outputs),
            initial learning rate,
            training batch size,
            number of boundary points,
            value and index of kinematic constant $c_{\nikv}$,
            kinematic scale bound (if applicable),
            physical cut size,
            and spurious cut size.
        The same hyperparameters are used to train \acp{NN} for both the real and imaginary parts of the solutions.
    }
    \label{tab:hyperparameters}
\end{table}

\begin{table}
    \centering
    \begin{tabular}{ c c c c }
        \hline
        Integral family & Final loss & Iterations & Time (minutes) \\
        \hline
        box & $2.7\times 10^{-7}$ & $2.5 \times 10^5$ & 16 \\
        one-mass double box & $3.4\times 10^{-4}$ & $1.1 \times 10^5$ & 53 \\
        heavy crossed box & $1.4\times10^{-5}$ & $7.9\times10^4$ & 75 \\
        top double box & $7.1\times10^{-4}$ & $5.2\times10^{4}$ & 32 \\
        \hline
    \end{tabular}
    \caption{
        Summary of training statistics for all examples, showing
            mean loss value of final epoch,
            mean total number of iterations,
            and typical time to train a single \ac{NN}.
        These results are averaged over all replicas within the ensemble.
    }
    \label{tab:training}
\end{table}

\begin{table}
    \centering
    \begin{tabular}{ c c c c c c c }
        \hline
        Integral family & MEU & MDE & MAD & MMRD & MLR & Size \\
        \hline
        box & $2.8\times 10^{-5}$ & $3.6\times10^{-4}$ & $2.9\times 10^{-5}$ & $2.2\times 10^{-5}$ & $3.9\times 10^{-7}$ & $10^5$ \\
        one-mass DB & $8.1\times 10^{-4}$ & $1.1\times10^{-2}$ & $2.0\times 10^{-3}$ & $1.1\times 10^{-2}$ & $-2.8\times 10^{-4}$ & $10^5$ \\
        heavy CB & $2.8\times10^{-4}$ & $2.8\times10^{-3}$ & $1.6\times10^{-3}$ & $7.3\times10^{-3}$ & $-4.5\times10^{-4}$ & $10^2$ \\
        top DB & $1.9\times10^{-4}$ & $1.7\times10^{-3}$ & $9.0\times10^{-4}$ & $3.9\times10^{-3}$ & $1.8\times10^{-4}$ & $10^2$ \\
        \hline
    \end{tabular}
    \caption{
        Summary of ensemble uncertainty and testing errors as defined in \cref{sec:error} for all examples.
        We use the shorthands DB and CB for double box and crossed box, respectively. 
        We show
            mean ensemble uncertainty (MEU),
            mean differential error (MDE),
            mean absolute difference (MAD),
            mean magnitude of relative difference (MMRD),
            mean logarithm of ratio (MLR),
            and size of testing dataset.
        All means are taken over all nonzero outputs of the ensemble, with the exception of the MDE, which is over all nonzero outputs of all replicas.
    }
    \label{tab:testing}
\end{table}

The first general feature we observe is related to the flatness of the performance of our \acp{NN} with respect to the analytic complexity of the learnt functions. 
From the analytic point of view, the real part of the integrals is more complicated than the imaginary part, the complexity grows with the depth in the $\eps$ expansion, and certain \acp{MI} are more complicated than others.
Our models instead appear to perform equally well for both complex parts, all orders in $\eps$, and all \acp{MI} of each family. 

Similarly, the performance appears to be flat also with respect to the complexity of the example.
In other words, after limited hyperparameter tuning, our models reach a similar level of performance for all the two-loop examples we considered.
We achieve a mean magnitude of relative difference of roughly 1\% for all the two-loop examples considered (see \cref{tab:testing}), with training times of the order of one hour on a laptop GPU (see \cref{tab:training}).
This is particularly promising, as it indicates that this approach is insensitive to the analytic complexity of the special functions appearing in the solution to the \acp{DE}. 
We recall in fact that, while for the one-mass double box integrals a convenient representation in terms of \acp{MPL} is well known, we have a representation in terms of Chen iterated integrals for the top double box integrals (which is not suitable for a fast numerical evaluation in the physical scattering region), and no analytic expression at all for the heavy crossed box.
It is important to emphasise however that the two-loop integral families we considered all have the same number of input variables ($2$, after one kinematical variable is fixed to a constant).
This warrants the analysis of Feynman integrals depending on more variables, which we defer to future studies.

Another promising observation which calls for a more in-depth study concerns the spurious singularities. 
As discussed in \cref{sec:data}, we impose cuts on the \ac{DE} dataset around the spurious singularities.
These cut regions are instead retained in the boundary and in the testing datasets.
In the examples we studied, this is relevant only for the top double box, as in the other cases the spurious singularities fall outside the physical scattering region under consideration.
Nonetheless, it is interesting that the extra cut on the \ac{DE} dataset does not appear to impact the performance of the model for the top double box even in proximity of the spurious singularity.
Feynman integrals with more intricate spurious singularities within the physical scattering region of interest should be analysed in the future to confirm this promising observation.
We recall that we instead treat the cuts on the physical singularities as hard cuts.
In other words, we did not verify whether the models extrapolate beyond them.
While it would be interesting to study this aspect as well, setting the physical cuts from the start so that they match the cuts in the intended application would ensure an optimal performance.

Finally, any numerical value is of little use without a reliable estimate of its uncertainty.
In particular, we need a prescription to associate the prediction of our model with an uncertainty without relying on the comparison with a large testing dataset.
In this work, we consider two quantities which may give such an estimate: the mean ensemble uncertainty and the mean differential error (see \cref{sec:error}).
Both of them are absolute measures, and we find them to be correlated to the mean absolute difference in the testing dataset.
Associating a reliable uncertainty to the model predictions would however require a more thorough analysis of the correlation between the ensemble uncertainty/differential error and the testing errors, which we leave for future study.

\section{Conclusion and outlook}
\label{sec:conc}

In this paper, we propose a new approach to evaluate Feynman integrals numerically.
We view the Feynman integrals as the solutions to systems of first-order \acp{PDE}.
We compute the \acp{PDE} analytically, and train a deep \ac{NN} to approximate their solution using the recently-proposed framework of \acf{PIDL}.
The \ac{NN} takes the kinematic variables as inputs, and returns the values of the integrals expanded around $\eps=0$ up to a chosen order, split into real and imaginary parts.
The method relies on the analytic expression of the \acp{DE}, and on the values of the Feynman integrals at a small number of reference points.
Mild assumptions are made on the form of the \acp{DE}, merely to simplify the problem, but we do not rely on a canonical form of the \acp{DE}.
Indeed, our method is aimed in particular at those cases where canonical \acp{DE} have not been obtained, or where their analytical solutions contain classes of special functions for which the mathematical technology is not yet fully developed.

We provide a proof-of-concept implementation within the \texttt{PyTorch} framework, and apply it to a number of one- and two-loop integral families.
One family is at the cutting edge of current analytical methods, and another is not yet known analytically.
In all two-loop examples, we achieve a mean magnitude of relative difference of order 1\% in the physical phase space with \ac{NN} training times on the order of an hour on a laptop GPU.
The performance appears to be flat with respect to the analytic complexity of the solutions.
In other words, after limited hyperparameter tuning, our models reach a similar level of performance for all examples we considered and, within a single family, for all integrals and all orders in their $\eps$ expansion.

We emphasise that the examples considered in \cref{sec:Examples} are only meant to prove the feasibility of our new approach.
Even the cases where no analytic solution is known may in fact be tackled with other standard numerical methods.
The top double box family of \cref{sec:topbox}, for example, was computed in~\cite{Czakon:2013goa,Barnreuther:2013qvf} by solving numerically the \acp{DE} at $\sim 1000$ phase-space points, and using the results to set up an interpolation grid. 
Our method, on the other hand, requires as input the values of the integrals at substantially fewer phase-space points.
This is therefore promising in view of the application to Feynman integrals for higher-multiplicity processes. 
The higher complexity of the phase space makes the construction of interpolation grids less effective, as the number of grid points scales badly with the number of variables, and  strengthens the demand for a fast numerical evaluation, as the number of points required in the Monte Carlo integration grows quickly as well.

With respect to traditional numerical \ac{DE} solver, semi-analytical \ac{DE} solvers based on generalised power series expansions, and Monte Carlo integration methods, our approach yields essentially instantaneous evaluation times.
The overall time budget is concentrated in the training of the \acp{NN}, which however is done once and for all for a given integral family.
This however comes at the price of a lower control over the uncertainty. 
We proposed a number of proxies to estimate the latter, based on the standard error of an ensemble of \acp{NN} fitting the solution to the same \acp{DE}, and on how well the model satisfies the \acp{DE} (differential error, see \cref{eq:diff-err}). 
Our preliminary investigation indicates that these proxies may be used to provide a reliable estimate of the uncertainty, although further study is required in this direction, in particular to characterise the correlation between the differential error and the testing error. 
However, controlling the accuracy, namely constructing and training a \ac{NN} such that it reaches a given target accuracy, remains an open problem.
This is particularly important for using the integrals in the computation of scattering amplitudes, where large cancellations among different terms may lead to a loss of accuracy.

The question of how to improve the fitting performance of \ac{PIDL} models is a well-known open challenge~\cite{hao2023physicsinformed}, although a number of refinements are being explored in the literature~\cite{wang2023experts}.
A hard limit on the achievable performance is set by the floating-point arithmetic. 
We have verified that we have not yet reached this limit by observing that training the models with double-precision arithmetic does not improve their performance.
One way to improve consists in performing further hyper-parameter tuning. 
For example, increasing the width and depth of the neural networks may enhance their expressivity, while changes in the optimisation procedure may allow us to find lower minima of the loss function. 
We stress that, in this study, we have performed limited hyper-parameter tuning, stopping when we deemed that the achieved performance was sufficient to prove the feasibility of our method.
An important feature of our target functions is that they exhibit
varied frequencies over a large distribution of scales. 
This can degrade stochastic gradient descent by introducing unbalanced back-propagated gradients~\cite{wang2020understanding,wang2020pinns}, while \acp{NN} are biased towards learning global fluctuations over local details~\cite{pmlr-v97-rahaman19a,ijcai2021p304}.
Of particular promise for our application are attempts to balance the training algorithm in order to address these issues, such as adaptive sampling of the phase space~\cite{Nabian_2021,daw2023mitigating,Wu_2023} or adaptive reweighting of the loss function~\cite{wang2020understanding,wang2020pinns,MCCLENNY2023111722,Maddu_2022}.

In addition to these developments on the machine learning side, there is room for improvement also on the physics side. 
Even when the integrals cannot be computed analytically, a lot is known about their asymptotic behaviour in the singular limits, either from the \acp{DE} themselves or through other approaches such as the method of regions~\cite{Beneke:1997zp,Smirnov:1999bza,Jantzen:2011nz}.
This precious information may be used for instance to partition the phase space based on the infrared singularities and then train a different \ac{NN} in each sub-region~\cite{Badger:2020uow,Aylett-Bullock:2021hmo}. 
Furthermore, one could impose hard constraints in the form of an ansatz with \acp{NN} as coefficients, along the lines of the factorisation-aware matrix-element emulation proposed in \incites{Maitre:2021uaa,Janssen:2023ahv,Maitre:2023dqz}.

Given the successful proof of concept discussed in this paper and the large room for improvement, we believe that this novel method has the potential to increase the evaluation efficiency of Feynman integrals for which the analytic solution is still out of reach.

\acknowledgments

We are indebted to Simon Badger, Christian Biello and Johannes Henn for many insightful discussions and comments on the manuscripts.
We further thank Matteo Becchetti and Ekta Chaubey for useful correspondence regarding the results of \incites{Adams:2018kez,Adams:2018bsn} and \incite{Becchetti:2023wev}, respectively, Tiziano Peraro for providing the \ac{MI} basis for the example in \cref{sec:topbox}, and Christoph Dlapa and Juan M.\ Cruz-Martinez for useful discussions.
F.C.\ wishes to thank the Max Planck Institute for Physics for hospitality and support, as well as the TMP Master program.
This project received funding from the European Union's Horizon 2020 Research and Innovation Programme \textit{High precision multi-jet dynamics at the LHC} (grant agreement No.~772099), and from the European Union's Horizon Europe Research and Innovation Programme under the Marie Skłodowska-Curie grant agreement No.~101105486.

\appendix

\section{Integral family definitions}
\label{app:families}

In this appendix, we collect the definitions of the propagators of the two-loop four-point integral families studied in \cref{sec:Examples}.
They all have 7 propagators, followed by 2 auxiliary propagators (called as \acp{ISP}),
which are required in order to accommodate into the integral family any numerator over the given propagator structure.
The integrals of each family have the form
\begin{align}
    \mathrm{I}_{\vec{a}}(\vec{v}; \eps) = \int \frac{\dd^d k_1}{\i \pi^{d/2}}  \frac{\dd^d k_2}{\i \pi^{d/2}}  \frac{ \mu^{2(4-d)} }{D_1^{a_1} D_2^{a_2} \cdots D_9^{a_9}} \,,
\end{align}
where $\vec{v}$ are the independent kinematic variables, $\mu$ is the dimensional regularisation scale, $\vec{a} = (a_1,\ldots,a_9) \in \mathbb{Z}^9$, and $D_i$ are the inverse propagators.
The last two are \acp{ISP}, hence $a_8, a_9 \le 0$.
We set $\mu = 1$.
We define below the inverse propagators $D_i$ family by family, omitting Feynman's $\i 0^+$ prescription.

\paragraph{One-mass double box (\cref{fig:1m-doublebox})}

\begin{align} \begin{aligned}
    \{D_i\}_{i=1}^{9} = \bigl\{ & k_1^2, \, (k_1 - p_1)^2, \, (k_1 - p_1 - p_2)^2, \, k_2^2, \, (k_2 - p_4)^2 , \, \\
    & (k_2 + p_1 + p_2)^2, \, (k_1 + k_2)^2, \, (k_1 + p_4)^2 , \, (k_2 + p_1)^2 \bigr\} \,.
\end{aligned} \end{align}

\paragraph{Heavy crossed box (\cref{fig:heavycrossedbox})}

\begin{align} \begin{aligned}
    \{D_i\}_{i=1}^{9} = \bigl\{ & k_1^2, \, (k_1 - p_1)^2, \, (k_1 + p_2)^2, \, k_2^2 - m^2, \, (k_1 + k_2 - p_1)^2 - m^2 , \\
    & (k_2 - p_4)^2 - m^2 , \, (k_1 + k_2 - p_1 + p_3)^2 - m^2, \, (k_1 + p_3)^2, \, (k_1 + k_2)^2 \} \,.
\end{aligned} \end{align}

\paragraph{Top double box (\cref{fig:topbox})}

\begin{align} \begin{aligned}
    \{D_i\}_{i=1}^{9} = \bigl\{ &  k_1^2 - m_{\rm t}^2, \, (k_1 - p_1)^2 - m_{\rm t}^2 , \, (k_1 - p_1 - p_2)^2 - m_{\rm t}^2 , \,
    k_2^2, \,  (k_2 - p_4)^2 -m_{\rm t}^2 \\
    & (k_2 - p_3 - p_4)^2, \, (k_1 + k_2)^2 - m_{\rm t}^2, \, (k_1 + p_4)^2, \, (k_2 + p_1)^2
    \bigr\} \,.
\end{aligned} \end{align}

\section{Supplementary plots}
\label{app:plots}

In this appendix we gather sets of plots to portray the training and testing statistics for each two-loop integral family analysed in this work. 
Each set mirrors \cref{fig:box-stats} for the one-loop massless box.
Since the two-loop families we consider depend on two variables, we trade the histogram in \cref{fig:err-ps-box} for a 2d histogram in the case of the one-mass double box ---~for which we have an analytic solution~--- or a 2d scatterplot in the other cases.
Furthermore, to keep this plot easy to read, we only show the absolute testing error and disregard the ensemble uncertainty.
We recall that the latter is in fact expected to catch only part of the uncertainty, and further study is required to define a robust uncertainty estimate (see \cref{sec:GeneralComments}).
We repeat some plots already presented in the main text to facilitate the comparison.

\newpage

\subsection{One-mass double box}
\label{app:t331ZZZM}
\begin{figure}[h]
    \centering
    
      \begin{subfigure}[c]{0.49\linewidth}
        \centering
        \includegraphics[width=\textwidth]{training-t331ZZZM}
        \caption{Learning curve.}
        \label{fig:training-t331ZZZM-app}
    \end{subfigure}
    \begin{subfigure}[c]{0.49\linewidth}
        \centering
        \includegraphics[width=\textwidth]{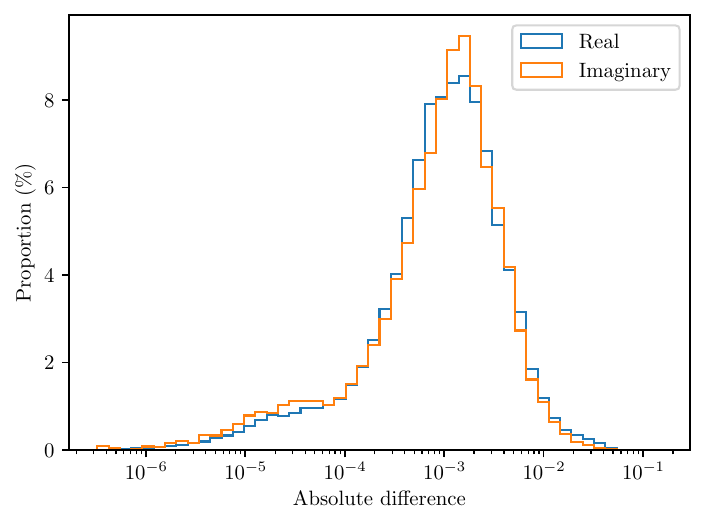}
        \caption{Absolute difference (real vs.\ imaginary).}
        \label{fig:abs-diff-part-t331ZZZM-app}
    \end{subfigure}
    \begin{subfigure}[c]{0.49\linewidth}
        \centering
        \includegraphics[width=\textwidth]{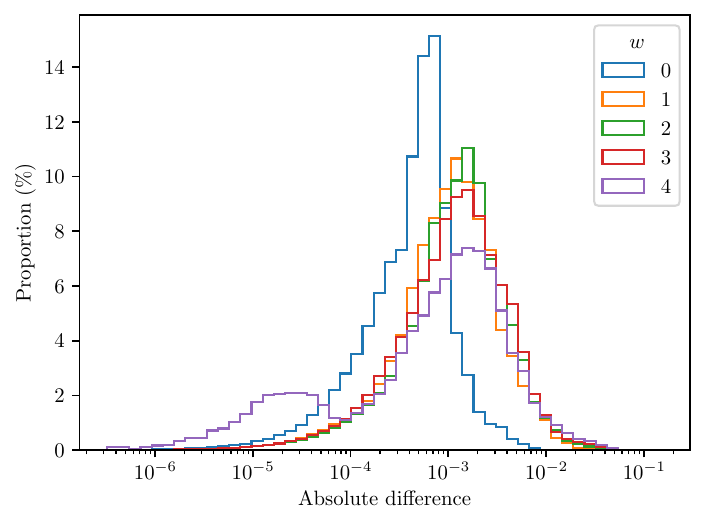}
        \caption{Absolute difference ($\eps$ orders).}
        \label{fig:abs-diff-eps-t331ZZZM-app}
    \end{subfigure}
    \begin{subfigure}[c]{0.49\linewidth}
        \centering
        \includegraphics[width=\textwidth]{rel_diff_eps_t331ZZZM}
        \caption{Magnitude of relative difference ($\eps$ orders).}
        \label{fig:rel-diff-t331ZZZM-app}
    \end{subfigure}
    \begin{subfigure}[c]{0.49\linewidth}
        \centering
        \includegraphics[width=\textwidth]{ratio_eps_t331ZZZM}
        \caption{Logarithm of ratio ($\eps$ orders).}
        \label{fig:ratio-t331ZZZM-app}
    \end{subfigure}
     \begin{subfigure}[c]{0.49\linewidth}
        \centering
        \includegraphics[width=\textwidth]{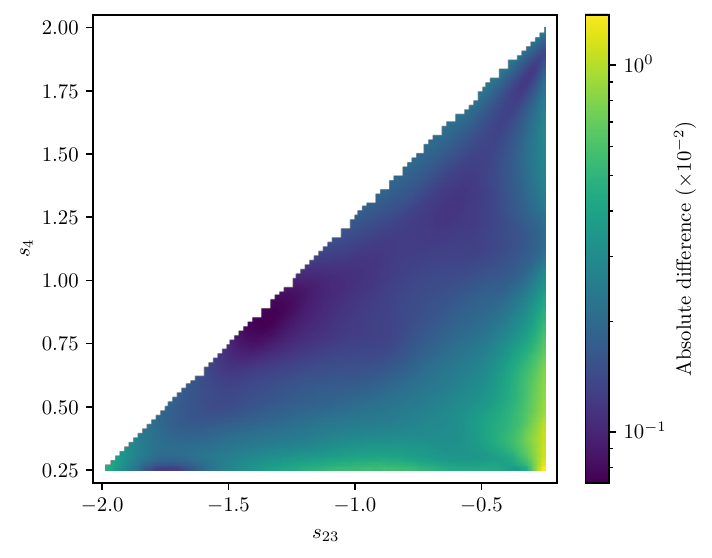}
        \caption{2d histogram of the absolute testing error.}
        \label{fig:histo-t331ZZZM-app}
    \end{subfigure}

    \caption{
        Full training and testing statistics for the one-mass double box discussed in \cref{sec:1m-doublebox}.
    }
    \label{fig:dblbox-stats}
\end{figure}

\newpage
\subsection{Heavy crossed box}
\label{app:heavycrossbox}
\begin{figure}[h]
    \centering
    
      \begin{subfigure}[c]{0.49\linewidth}
        \centering
        \includegraphics[width=\textwidth]{training-heavycrossbox}
        \caption{Learning curve.}
        \label{fig:training-heavycrossbox-app}
    \end{subfigure}
    \begin{subfigure}[c]{0.49\linewidth}
        \centering
        \includegraphics[width=\textwidth]{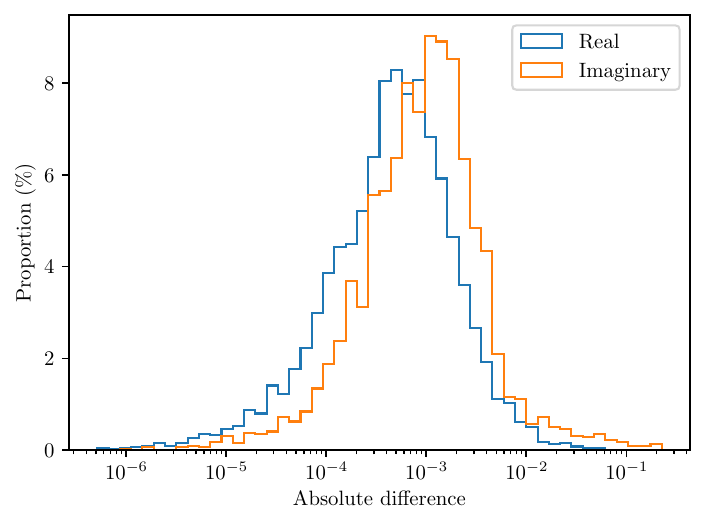}
        \caption{Absolute difference (real vs.\ imaginary).}
        \label{fig:abs-diff-part-heavycrossbox-app}
    \end{subfigure}
    \begin{subfigure}[c]{0.49\linewidth}
        \centering
        \includegraphics[width=\textwidth]{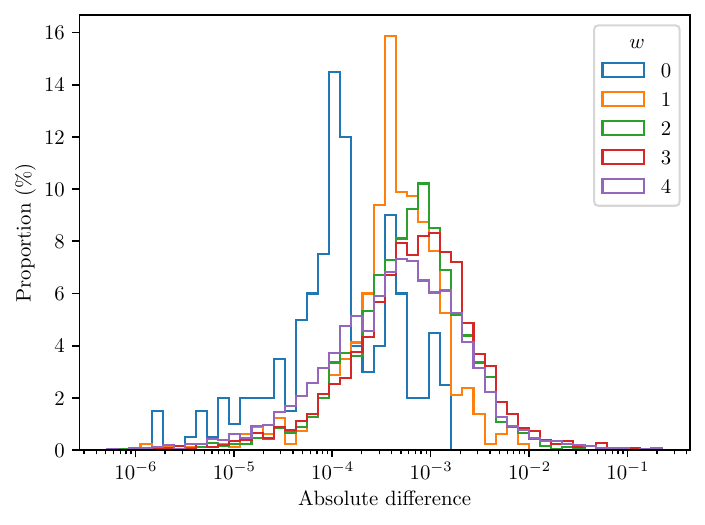}
        \caption{Absolute difference ($\eps$ orders).}
        \label{fig:abs-diff-eps-heavycrossbox-app}
    \end{subfigure}
    \begin{subfigure}[c]{0.49\linewidth}
        \centering
        \includegraphics[width=\textwidth]{rel_diff_eps_heavycrossbox}
        \caption{Magnitude of relative difference ($\eps$ orders).}
        \label{fig:rel-diff-heavycrossbox-app}
    \end{subfigure}
    \begin{subfigure}[c]{0.49\linewidth}
        \centering
        \includegraphics[width=\textwidth]{ratio_eps_heavycrossbox}
        \caption{Logarithm of ratio ($\eps$ orders).}
        \label{fig:ratio-heavycrossbox-app}
    \end{subfigure}
     \begin{subfigure}[c]{0.49\linewidth}
        \centering
        \includegraphics[width=\textwidth]{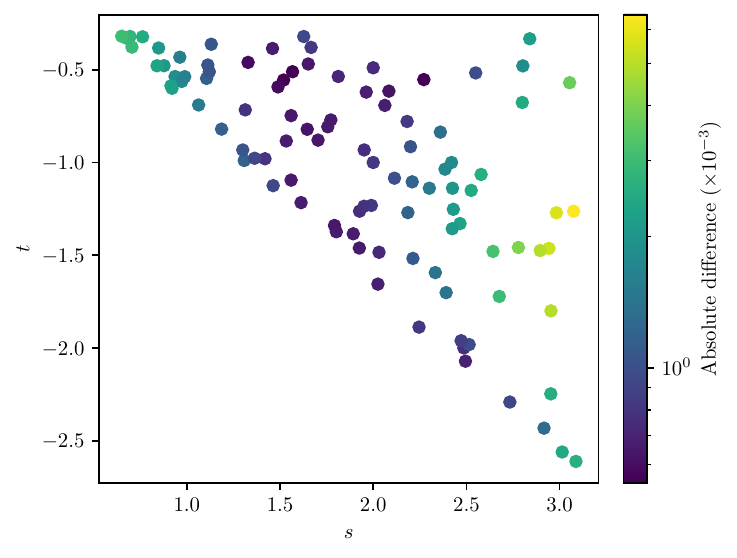}
        \caption{2d scatterplot of the absolute testing error.}
        \label{fig:histo-heavycrossbox-app}
    \end{subfigure}

    \caption{
        Full training and testing statistics for the heavy crossed box discussed in \cref{sec:heavycrossedbox}.
    }
    \label{fig:hcb-stats}
\end{figure}

\newpage
\subsection{Top double box}
\label{app:topbox}
\begin{figure}[h]
    \centering
    
      \begin{subfigure}[c]{0.49\linewidth}
        \centering
        \includegraphics[width=\textwidth]{training-topbox}
        \caption{Learning curve.}
        \label{fig:training-topbox-app}
    \end{subfigure}
    \begin{subfigure}[c]{0.49\linewidth}
        \centering
        \includegraphics[width=\textwidth]{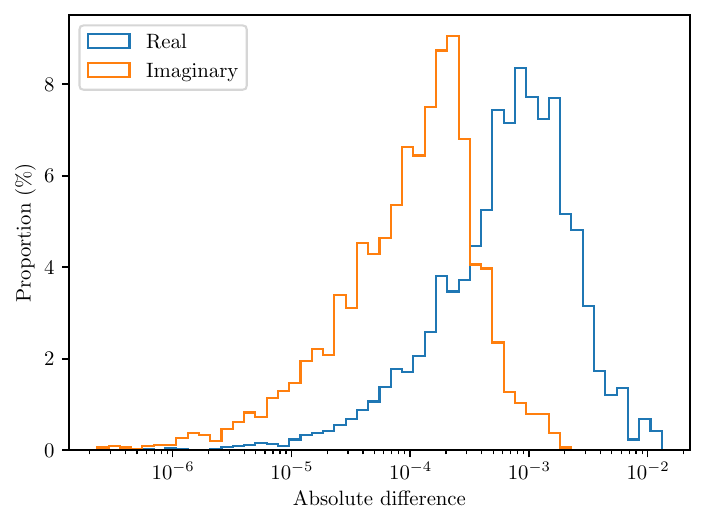}
        \caption{Absolute difference (real vs.\ imaginary).}
        \label{fig:abs-diff-part-topbox-app}
    \end{subfigure}
    \begin{subfigure}[c]{0.49\linewidth}
        \centering
        \includegraphics[width=\textwidth]{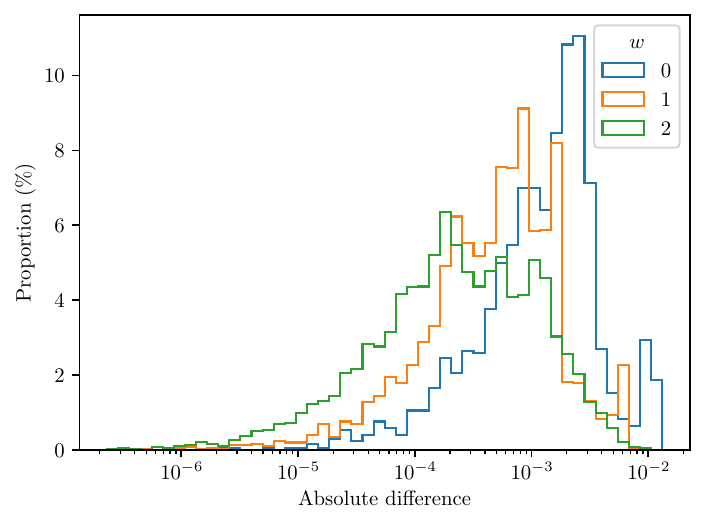}
        \caption{Absolute difference ($\eps$ orders).}
        \label{fig:abs-diff-eps-topbox-app}
    \end{subfigure}
    \begin{subfigure}[c]{0.49\linewidth}
        \centering
        \includegraphics[width=\textwidth]{rel_diff_eps_topbox}
        \caption{Magnitude of relative difference ($\eps$ orders).}
        \label{fig:rel-diff-topbox-app}
    \end{subfigure}
    \begin{subfigure}[c]{0.49\linewidth}
        \centering
        \includegraphics[width=\textwidth]{ratio_eps_topbox}
        \caption{Logarithm of ratio ($\eps$ orders).}
        \label{fig:ratio-topbox-app}
    \end{subfigure}
     \begin{subfigure}[c]{0.49\linewidth}
        \centering
        \includegraphics[width=\textwidth]{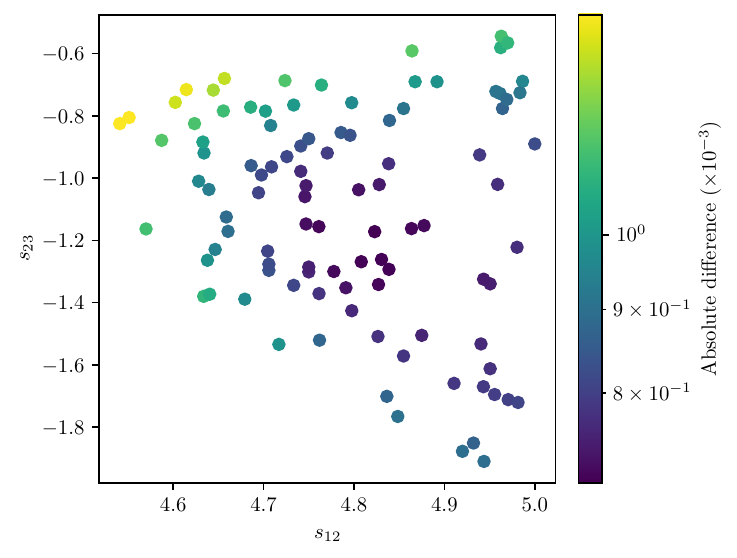}
        \caption{2d scatterplot of the absolute testing error.}
        \label{fig:histo-topbox-app}
    \end{subfigure}

    \caption{
        Full training and testing statistics for the top double box discussed in \cref{sec:topbox}.
    }
    \label{fig:tb-stats}
\end{figure}

\bibliographystyle{JHEP}
\bibliography{bibliography}

\end{document}